\documentclass[svgnames]{article}

\usepackage{authblk}

\newcommand*{\email}[1]{%
    \par\normalsize\href{mailto:#1}{#1}\par
}

\usepackage{booktabs} 
\usepackage{wrapfig}
\usepackage{array}

\newcommand*{\vertbar}{\rule[-1ex]{0.5pt}{2.5ex}}
\newcommand*{\horzbar}{\rule[.5ex]{2.5ex}{0.5pt}}

\usepackage[linesnumbered,ruled]{algorithm2e}
\usepackage{forest}
\usepackage{amsmath}
\usepackage{amssymb}
\usepackage{amsthm}
\usepackage{balance}
\usepackage[inline]{enumitem}
\usepackage{subcaption}
\usepackage{adjustbox}
\usepackage{pdflscape}
\usepackage{lscape}
\usepackage{csquotes}
\usepackage{xcolor}
\usepackage{hyperref}
\usepackage{doi}
\usepackage{orcidlink}

\usepackage{pgfplots}
\usepackage{pgfplotstable}

\usepgfplotslibrary{units}
\usepgfplotslibrary{colorbrewer}
\usepgfplotslibrary{groupplots}
\usetikzlibrary{patterns}

\pgfplotsset{
    compat=newest,
    cycle list/Set1,
    scaled x ticks=false,
    scaled y ticks=false
}

\usepackage{todonotes}

\newcommand*{\drawCircle}[1]{%
    \raisebox{-.5ex}{\tikz{\draw[fill=#1,line width=.5pt] circle[radius=1ex];}}%
}%

\tikzset{
  labeled node/.style={
    rectangle split, rectangle split parts=2,
    rectangle split draw splits=false,
    rounded corners,
    very thick,
    draw=blue!50,
    rectangle split part fill={blue!50, white},
    align=left,
    font=\sffamily\bfseries\boldmath\small
  },
  labeled edge/.style={
    rectangle split, rectangle split parts=1,
    rectangle split draw splits=false,
    rounded corners,
    very thick,
    draw=orange!60,
    rectangle split part fill={orange!60, white},
    align=left,
    font=\sffamily\bfseries\boldmath\small
  },
  labeled property edge/.style={
    rectangle split, rectangle split parts=2,
    rectangle split draw splits=false,
    rounded corners,
    very thick,
    draw=orange!60,
    rectangle split part fill={orange!60, white},
    align=left,
    font=\sffamily\bfseries\boldmath\small
  },
  every two node part/.style={font=\sffamily\small},
}

\newtheorem{definition}{Definition}
\newtheorem{proposition}{Proposition}
\newtheorem{example}{Example}
\newtheorem{axiom}{Axiom}

\def\architxt{ArchiTXT}

\begin{document}
\title{From Text to Databases: attribute grammar as database meta-model}
  
\author[1]{Jacques Chabin\orcidlink{0000-0003-1460-9979}} 
\author[1]{Mirian Halfeld-Ferrari\orcidlink{0000-0003-2601-3224}}
\author[1]{Nicolas Hiot\orcidlink{0000-0003-4318-4906}}

\affil[1]{\small Universit\'e d'Orl\'eans, INSA CVL, LIFO,  UR 4022, Orléans, France.
\email{\{jacques.chabin, mirian, nicolas.hiot\}@univ-orleans.fr}}

\date{October 17, 2024}

\maketitle

\begin{abstract}
We present a general methodology for structuring textual data, represented as syntax trees enriched with semantic information, guided by a meta-model $\mathbb{G}$ defined as an attribute grammar. The method involves an evolution process where both the instance and its grammar evolve, with instance transformations guided by rewriting rules and a similarity measure. Each new instance generates a corresponding grammar, culminating in a target grammar $G_T$ that satisfies $\mathbb{G}$. 

This methodology is applied to build a database populated from textual data. The process generates both a database schema and its instance, independent of specific database models. We demonstrate the approach using clinical medical cases, where trees represent database instances and grammars act as database schemas. Key contributions include the proposal of a general attribute grammar $\mathbb{G}$, a formalization of grammar evolution, and a proof-of-concept implementation for database structuring.
\end{abstract}

\section{Introduction}
\label{sec:Intro}


In this work we are interested in structuring textual data, represented as a rooted forest (i.e., syntax trees combined into a single structure with a common root), through a method based on the idea that texts follow an initial grammar, $G_0$. These trees can be unified into a single structure with a common root. The grammar $G_0$ serves as a schema for the initial text. Structuring involves transforming this instance to satisfy the constraints of a new schema defined by a target grammar, $G_T$. A meta-model, $\mathbb{G}$, guides the process by outlining the desired structure, with transformations applied to achieve it. This process unfolds as a sequence of steps that evolve both the instance and the grammar from $G_0$ to $G_T$. The method, which is based on structural transformation, can be enhanced by enriching the syntax trees with pre-processed knowledge, such as detected named entities, or by incorporating semantic constraints in $\mathbb{G}$.

Our process involves the generation of both the database schema and its corresponding instance. Rather than targeting a specific database model, we adopt a general abstraction that captures common concepts and relationships found in database design, independent of the implementation.

Recent years have seen an explosion in the production of unstructured data, particularly textual data.
On the one hand, this plethora of text data provides valuable information for research and decision-making.
Processing this data is important for extracting insights, identifying trends, automating tasks and making informed decisions. This poses a substantial challenge for AI and data analytics:
uncovering insights in unstructured data  requires complex analytics and advanced technical skills.
Structured data, on the other hand, with its well-defined schema and relationships, enables efficient analysis and database storage. It makes it easier to check constraints and allows reuse by different users.

When it comes to considering how to organize information that originally comes from a text, a critical consideration is the choice of an appropriate data model. Currently, the primary options typically range between relational databases, which excel in maintaining data consistency and integrity despite challenges related to schema evolution, and NoSQL models such as key-value stores, document stores, or graph databases, which are favored for their flexibility and scalability.
However, complicating matters further, certain applications require interfacing with multiple data models simultaneously. In such scenarios, a meta data model is regarded as a viable solution to facilitate interchange and integration across different models. This meta model enables seamless translation between disparate data structures, thereby enhancing interoperability and system functionality.

Our method deals with the structuring of textual data to populate a database.
We developed our approach using clinical medical cases as the application domain.  Starting from textual descriptions, we generate a generic hierarchical structure representing a database schema and its instance.  This instance consists of key text fragments suitable for database storage. Our approach is hybrid: it relies on syntax to transform the initial tree by pruning or aggregating sub-trees, but begins with an enriched tree containing semantic information extracted during pre-processing, taking into account the application domain. These transformations are thus indirectly guided by this enrichment step.
Thus, in our approach, trees represent the  database instance,  and a grammar capable of generating such trees serves as  the database schema.

\vspace{0.3cm}
\noindent
In this context, the main contributions of this paper are:

\begin{itemize}[leftmargin=*]
\item The proposal of an original way of looking at the structuring of textual data.
    \item The proposal of an attribute grammar $\mathbb{G}$ that represents a  generic database structure.
          A  grammar respecting $\mathbb{G}$ is seen as a database schema which can be then translated to any  data model such as relational or graph models.

    \item The formalization of textual data structuring through the evolution process of a grammar $G_0$ (associated to the syntactic trees of the sentences) to a target grammar $G_T$ which respects  the meta-grammar $\mathbb{G}$.

    \item The formalization of an evolution process through transformations on the trees (instances) guided by tree rewriting rules and a similarity measure.
    
    \item A proof-of-concept implementation  that uses clinical cases as input and $\mathbb{G}$ as a generic database structure.
   \end{itemize}

\noindent
\textbf{Paper Organization.} Section~\ref{sec:Over}  overviews our approach.
In Section~\ref{sec:RelW} we position our work with respect to some related work.
Section~\ref{sec:Back} gives some background concepts.
Section~\ref{sec:autoStruc} presents the details of our structuring method, including the definition of our meta-grammar $\mathbb{G}$ (Section~\ref{sec:MG}). 
Section~\ref{sec:POC} carefully analyses a proof of concept and Section~\ref{sec:conclusion} offers some final comments.

\section{Overview}
\label{sec:Over}
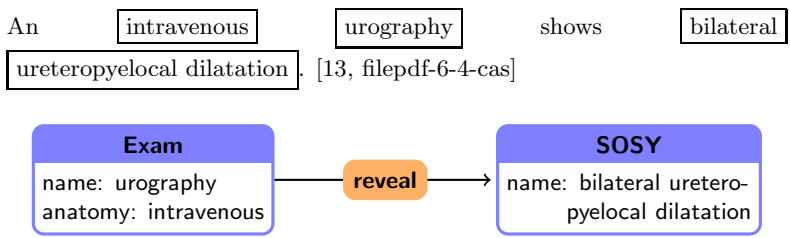
\begin{figure}[htb]
  \centering
  \begin{displaycquote}[filepdf-6-4-cas]{grabarCASFrenchCorpus2018}\small
    An \fbox{intravenous} \fbox{urography} shows \fbox{bilateral} \fbox{ureteropyelocal dilatation}.
  \end{displaycquote}
  \vspace{.2em}
  \begin{adjustbox}{max width=\linewidth, max height=.15\textheight}
    \begin{tikzpicture}[shorten >=2pt,thick,->,node distance=5em and 18em,on grid]
      \node[labeled node] (exam) {Exam \nodepart{two} name: urography\\anatomy: intravenous};
      \node[labeled node, right=of exam] (sosy) {SOSY \nodepart{two} name: bilateral uretero-\\\hspace{.8cm}pyelocal dilatation};

      \path
      (exam) edge node[labeled edge, anchor=center] {reveal} (sosy)
      ;
    \end{tikzpicture}
  \end{adjustbox}
  \caption{An example of a graph database instance generated by structuring a text describing a clinical case.}
  \label{fig:exGeneral}
\end{figure}

The ultimate goal of our approach is to organize textual information by extracting a database instance from the text, as demonstrated in Figure~\ref{fig:exGeneral}. This structuring method involves altering the level of abstraction, allowing for the generalization of information when feasible.

\begin{figure}[htb]
  \sbox0{
    \begin{subfigure}{.55\linewidth}
      \begin{adjustbox}{max width=\linewidth, max height=.15\textheight}
        \begin{forest}
          for tree={s sep=.5em,l sep=1em,l=1em,fit=tight},
          where n children=0{tier=word}{}
          [S
            [NP
              [DT [The]]
              [NN [heart]]
              [NN [rate]]
            ]
            [VP
              [VBD [was]]
              [NP
                [CD [100]]
                [NN [bpm]]
              ]
            ]
          ]
        \end{forest}
      \end{adjustbox}
      \caption{Syntactic tree}
      \label{fig-trees-ex1-stx}
  \end{subfigure}}
  \sbox1{
    \begin{subfigure}{.55\linewidth}
      \centering
      \begin{adjustbox}{max width=\linewidth, max height=.13\textheight}
        \begin{forest}
          for tree={s sep=.5em,l sep=1em,l=1em,fit=tight},
          where n children=0{tier=word}{}
          [S
            [NP
              [DT [The]]
              [$ENT_{SOSY}$
                [NN [heart]]
                [NN [rate]]
              ]
            ]
            [VP
              [VBD [was]]
              [NP
                [CD [$ENT_{VALUE}$ [100]]]
                [NN [$ENT_{UNIT}$ [bpm]]]
              ]
            ]
          ]
        \end{forest}
      \end{adjustbox}
      \caption{Enriched tree}
      \label{fig-trees-ex1-ent}
  \end{subfigure}}
  \sbox2{
    \begin{subfigure}{.4\linewidth}
      \centering
      \begin{adjustbox}{max width=\linewidth, max height=.15\textheight}
        \begin{forest}
          for tree={s sep=.5em,l sep=1em,l=1em,fit=tight},
          where n children=0{tier=word}{}
          [S
            [NP
              [$ENT_{SOSY}$
                [NN [heart]]
                [NN [rate]]
              ]
            ]
            [VP
              [NP
                [CD [$ENT_{VALUE}$ [100]]]
                [NN [$ENT_{UNIT}$ [bpm]]]
              ]
            ]
          ]
        \end{forest}
      \end{adjustbox}
      \caption{Simplified tree}
      \label{fig-trees-ex1-simp}
  \end{subfigure}}
  \sbox3{
    \begin{subfigure}{.4\linewidth}
      \centering
      \begin{adjustbox}{max width=\linewidth, max height=.15\textheight}
        \begin{forest}
          for tree={s sep=.5em,l sep=1em,l=1em,fit=tight},
          where n children=0{tier=word}{}
          [S
            [$ENT_{SOSY}$
              [heart]
              [rate]
            ]
            [VP
              [$ENT_{VALUE}$ [100]]
              [$ENT_{UNIT}$ [bpm]]
            ]
          ]
        \end{forest}
      \end{adjustbox}
      \caption{Reduced tree}
      \label{fig-trees-ex1-reduce}
  \end{subfigure}}
  \centering
  \usebox0 \hfill \usebox2
  \usebox1 \hfill \usebox3
  \caption{Example of entity incorporation in a syntax tree and simplifications}
  \label{fig-trees-ex1}
\end{figure}
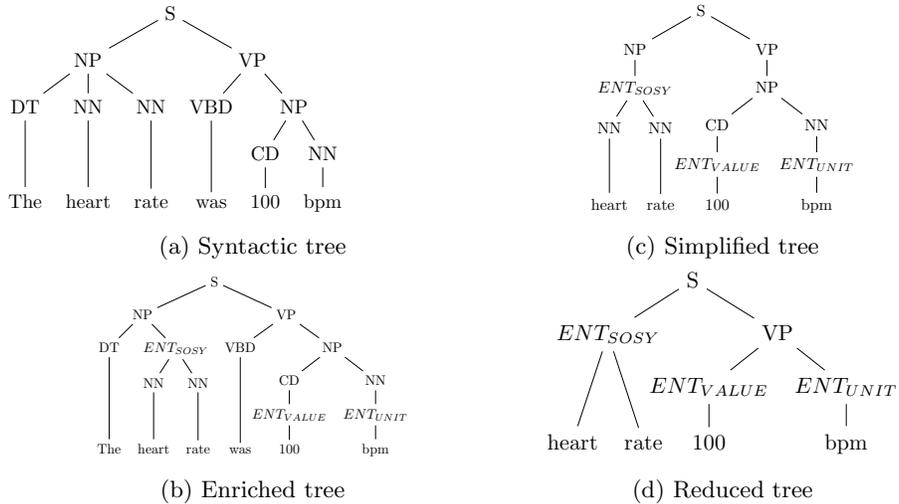

Our approach is grounded from a grammatical perspective: each sentence in the text is associated with its syntactic tree, and we consider the initial grammar $G_0$ as the one that accepts these syntactic trees.
Figure~\ref{fig-trees-ex1-stx} presents an example of a syntactic tree.

\begin{figure}[htb]
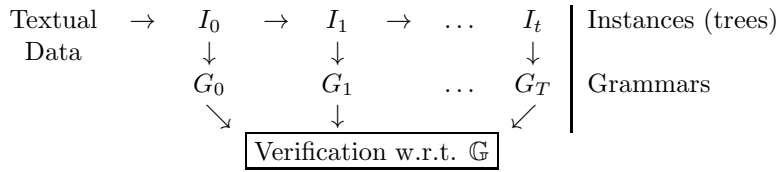

  \centering
  \begin{adjustbox}{max width=.9\linewidth}
    \begin{tabular}{cccccccc|l}
      Textual & $\rightarrow$ & $I_0$         & $\rightarrow$ & $ I_1$        & $\rightarrow$ & $ \dots$ & $I_t$         & Instances (trees) \\
      Data    &               & $\downarrow$  &               & $\downarrow$  &               &          & $\downarrow$  &                   \\
      &               & $G_0$         &               & $G_1$         &               & $ \dots$ & $G_T$         & Grammars          \\
      &               & ~ $\searrow$  &               & $\downarrow$  &               &          & $\swarrow$ ~  &                   \\
      &               & \multicolumn{6}{c}{\fbox{Verification w.r.t. $\mathbb{G}$}}                              &
    \end{tabular}
  \end{adjustbox}
  \caption{Iterative process for automatic structuring}
  \label{fig:struct:general-V0}
\end{figure}

Our proposal consists in  an iterative process which is visually summarized in Figure~\ref{fig:struct:general-V0}.
We work progressively by transforming the data instance (trees) and the data schema (a grammar).
Our goal is to obtain an instance respecting a target grammar $G_T$, which in turn respects the meta-grammar $\mathbb{G}$.

%
%
%

In Figure~\ref{fig:struct:general-V0}, the
vertical axis represents the extraction of a grammar from an instance $I_i$ in the form of a rooted forest of enriched syntactic trees, while
the horizontal axis  represents the progression of the process from step $i$ to step $i+1$ based on transformations on the instance.
These transformations follow  the following reasoning:

\begin{enumerate}[leftmargin=*]
  \item \textbf{Initialisation.} The process starts with the transformation of sentences into syntactic trees, which gives rise to an instance denoted by $I_0$.

  \item \textbf{Enrichment.}  The second  step is to enrich the trees of the $I_0$  instance by incorporating information previously extracted from the text analysis.
    This may involve inserting named entities, relationship or other relevant semantic information into the trees.

\item \textbf{Successive Evolutions:}
    \begin{enumerate}
      \item \textbf{Evolution of the instance}.
        To evolve from instance $I_i$ to instance $I_{i+1}$, the branches of the tree are grouped or transformed based on similarity measures. This may involve reorganising the structure of the sub-trees to improve their consistency or to better align them with the desired representation.

      \item \textbf{Evolution of the grammar}.
        The evolution of $G_i$ towards $G_{i+1}$ is triggered by checking whether the grammar $G_i$ conforms to the meta-grammar $\mathbb{G}$.
        At step $i$, if $G_i$ does not conform to $\mathbb{G}$, the process continues by transforming the tree structures of $I_i$, giving rise to the instance $I_{i+1}$, which generates 
        a new grammar $G_{i+1}$.
        The process ends when we find a grammar $G_T$ which satisfies $\mathbb{G}$.
    \end{enumerate}
\end{enumerate}

\begin{example}
  Figure~\ref{fig-trees-ex1} illustrates an example of transformation on a tree instance.
  In Figure~\ref{fig-trees-ex1-stx}, the syntactic tree corresponding to the sentence \enquote{The heart rate was 100 bpm} is shown.
  Figure~\ref{fig-trees-ex1-ent} displays the same tree enriched with the named entities $ENT_{SOSY}$, $ENT_{VALUE}$ and $ENT_{UNIT}$, identified during a pre-processing step.
  Figures~\ref{fig-trees-ex1-simp} et \ref{fig-trees-ex1-reduce} present trees where unnecessary terms and nodes have been removed.
  For instance, in Figure~\ref{fig-trees-ex1-simp}, only the branches corresponding to terms relevant to the database are retained (e.g., the branch corresponding to the article \enquote{The} in Figure~\ref{fig-trees-ex1-stx} is deleted).
  In Figure~\ref{fig-trees-ex1-reduce}, nodes indicating grammatical functions such as $VP$, $NN$, $NP$, $CD$, etc., are deemed irrelevant for the database context and are therefore removed.
  The objective here is to maintain the overall structure with the minimum number of nodes while preserving those that convey the necessary semantic information for database reasoning.\qed
\end{example}

Each iteration introduces modifications to the tree instance, guided by tree rewriting rules and a similarity measure that helps define equivalence classes.
The rewriting rules aims to prune sub-trees or reorganize them differently.
Section~\ref{sec:TransTrees} discusses this aspect in details.

\begin{figure}[htb]
  \begin{subfigure}{.3\linewidth}
    \centering
    \begin{adjustbox}{max width=\linewidth}
      \begin{forest}
        for tree={circle,draw,fit=tight,s sep=.8em,l=.2em,l sep=.5em}
        [, fill=white
          [, fill=red
            [, fill=black]
            [, fill=blue]
          ]
          [, fill=DarkGreen
            [, fill=blue]
            [, fill=red
              [, fill=black]
            ]
            [, fill=red
              [, fill=black]
            ]
          ]
        ]
      \end{forest}
    \end{adjustbox}
    \caption{Instance tree $I$}
    \label{fig-quotient-instance}
  \end{subfigure}
  \hfill
  \begin{subfigure}{.3\linewidth}
    \centering
    \begin{adjustbox}{max width=\linewidth}
      \begin{forest}
        for tree={circle,draw,fit=tight,s sep=.8em,l=.2em,l sep=.5em}
        [, fill=white,
          [, fill=red
            [, fill=black]
            [, fill=blue]
          ]
          [, fill=DarkGreen
            [, fill=blue]
            [, fill=red
              [, fill=black]
            ]
          ]
        ]
      \end{forest}
    \end{adjustbox}
    \caption{Quotient tree $Q$}
    \label{fig-quotient-tree}
  \end{subfigure}
  \hfill
  \begin{subfigure}{.3\linewidth}
    \centering
    \begin{align*}
      \drawCircle{white}     & \to \drawCircle{red}   ~ \drawCircle{DarkGreen} \\
      \drawCircle{red}       & \to \drawCircle{black} ~ \drawCircle{blue}      \\
      \drawCircle{DarkGreen} & \to \drawCircle{blue}  ~ \drawCircle{red}^+
    \end{align*}
    \caption{Grammar $G$}
    \label{fig-quotient-grammar}
  \end{subfigure}
  \caption{Example of quotient tree}
  \label{fig-quotient}
\end{figure}
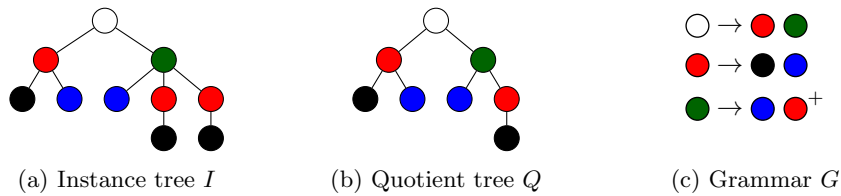

With reference to the vertical axis of Figure~\ref{fig:struct:general-V0}, at each step $i$, a grammar $G_i$ is derived from instance $I_i$ through the computation of a \textit{quotient tree}.
The idea behind this process is illustrated in Figure~\ref{fig-quotient} (see Section~\ref{sec:extractG} for formal definitions).

The instance tree in Figure~\ref{fig-quotient} is summarized into a quotient tree based on equivalence classes, highlighting the main patterns in the tree.
This quotient tree (Figure~\ref{fig-quotient-tree}) 
shows that the tree in Figure~\ref{fig-quotient-instance} has green nodes with blue and red children.
Once the quotient tree is created, extracting the corresponding grammar becomes straightforward, as shown in Figure~\ref{fig-quotient-grammar}.
In this example, each production rule in grammar $G$ associates a node's color with the color of its children.
For instance, a red node generates black and blue nodes, but in Figure~\ref{fig-quotient-instance}, some red nodes lack their blue child, indicating missing information.
Our approach accounts for such cases, enabling the extraction of grammars like $G$ in Figure~\ref{fig-quotient-grammar}.
Additionally, our grammars are expressed as extended context-free grammars, using symbols like $+$ to indicate repetition, as seen in the production rule for the green node.

\begin{figure}[htb]
  \centering
  \begin{subfigure}[c]{.4\linewidth}
    \centering
    \begin{adjustbox}{max width=\linewidth, max height=.15\textheight}
      \begin{forest}
        for tree={s sep=.5em,l sep=.5em,l=.5em,fit=tight},
        where n children=0{tier=word}{}
        [COLL$_1$
          [REL$_1$
            [GROUP$_1$
              [ENT$_1$ [$v_1$]]
              [ENT$_2$ [$v_2$]]
            ]
            [GROUP$_2$
              [ENT$_3$ [$v_3$]]
            ]
          ]
        ]
      \end{forest}
    \end{adjustbox}
  \end{subfigure}
  \hfill
  \begin{subfigure}[c]{.5\linewidth}
    \centering
    \small
    \begin{align*}
      \lambda & \to COLL_1            \\
      COLL_1  & \to REL_1^+           \\
      REL_1   & \to GROUP_1 ~ GROUP_2 \\
      GROUP_1 & \to ENT_1 ~ ENT_2     \\
      GROUP_2 & \to ENT_3
    \end{align*}
  \end{subfigure}

  \caption{An example of a tree instance after our iterative process, where internal nodes represent concepts from $\mathbb{G}$, along with its corresponding grammar $G_T$.}
  \label{fig:metaConcepts}
\end{figure}
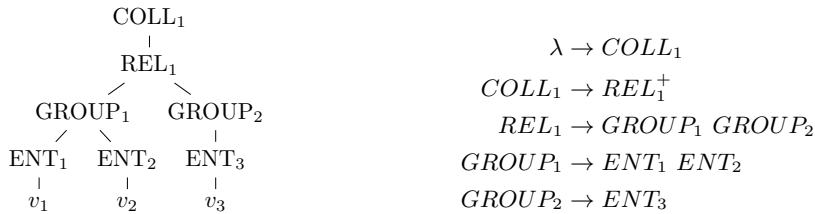

The target grammar $G_T$ is the one that respects the meta-grammar $\mathbb{G}$.
The purpose of $\mathbb{G}$ is to offer a general description of the main database abstractions, independent of the specific database model. 
$\mathbb{G}$ defines four key concepts that guide the generalizations and transformations applied to trees:
\begin{description}
  \item[Attribute] A name associated with a data value.
  \item[Group] A named set of attributes.
  \item[Relation] A relationship between distinct groups.
  \item[Collection] A set of equivalent groups and relations.
\end{description}
In Figure~\ref{fig:metaConcepts}, these concepts are illustrated within a tree instance.
The figure also displays the grammar extracted from this tree.

\section{Related Work}
\label{sec:RelW}
Text structuring can be considered through top-down and bottom-up perspectives \cite{abdullah2023systematic}.
In the top-down approach, a schema is provided, and the problem is seen as a query on the text to extract or identify \textit{relevant} information.
In \cite{jurafskySpeechLanguageProcessing2008,minardDOINGDEFTCascade2020,valenzuela2020odinson,savaryRelationExtractionClinical2022,cui2024semantic} we find examples of traditional approaches,
while recent trends are shifting towards machine learning (ML) and large language models (LLMs) for the extraction of entities and relationships \cite{li2020survey,keraghel2024survey,dagdelen2024structured}.
ML methods often depend on large annotated corpora for training, which can be both expensive and time-consuming. While they perform well, they may lack the flexibility to handle novel or unexpected data types that deviate from the established schema.


Bottom-up approaches  are  typically referred to as  \textit{open information extraction} (OpenIE). They are  used, for example, in  \textit{ontology learning} and involve the extraction of terms, entities, and relationships from text, followed by their classification and grouping, often based on similarity or syntactic rules \cite{liuDBpediaBasedEntityLinking2018,al-aswadiAutomaticOntologyConstruction2020}.
These methods operate without predefined schemas, offering greater flexibility across diverse domains.
Early techniques, such as those described in \cite{morinAutomaticAcquisitionSemantic1999,gamalloMappingSyntacticDependencies2002}, employed syntactic patterns to extract $\langle \text{subject}, \text{predicate}, \text{object} \rangle$ triples, linking them to knowledge bases.
Recent advancements also shifted towards ML techniques \cite{shenProbabilisticModelLinking2014,fanLinkingEntitiesRelations2024}, particularly leveraging neural networks to enhance extraction accuracy \cite{sakorFalconEntityRelation2020,liEfficientOnePassEndtoEnd2020,zhou2022survey,zhu2023large,navigliWordSenseDisambiguation2024}.
Even with these developments, ontology learning remains a complex challenge \cite{browarnikOntologyLearningText2015} that often requires human supervision, although LLMs tend to avoid it.
Current OpenIE models still struggle to extract meaningful relationships and lack a standardized output format~\cite{liu2022open}.

As mentioned in Section~\ref{sec:Intro} our approach is hybrid, combining syntax tree transformation with semantic enrichment of the tree beforehand. Hybrid methods have demonstrated potential for improved efficiency.
For instance, \cite{yu2021semi} introduces Semi-Open Information Extraction (SOIE) to discover domain-independent facts, and \cite{wilke2016merging} proposes a tool to merge a bottom-up graph from unstructured text with a top-down graph from structured data.

As data volume and variety grow, information system administrators must find effective solutions for storing, managing, and integrating data from multiple sources while addressing user needs. The concept of using a common meta-model that can be mapped to different database models is increasingly being proposed as a solution~\cite{barretAbstraGenericAbstractions2022,maliFACTDMFrameworkAutomated2024}.
In \cite{barretAbstraGenericAbstractions2022}, the authors propose schema extraction for structured and semi-structured data coming in different formats, with the aim of presenting datasets uniformly to help users understand and make choices. They use an intermediate structure based on concepts (sub-records, records, collections) similar to those in our generic schema (entities, groups, relations, collections). For this reason, we consider it the work most closely related to ours.
Unlike our method, which transforms syntax trees (text) to generate new instances, their approach does not handle unstructured data. Instead, they map (semi-)structured data from various sources to a graph format, detecting nodes that correspond to the key concepts they define. By interpreting these nodes, data from different sources can be better understood and integrated.
Their focus is on helping users understand the data; once nodes have been classified based on some user input, they are given semantic meaning, often using an ontology.
This last step is beyond the scope of our work, as our grammar is not designed for non-specialist users.

In~\cite{maliFACTDMFrameworkAutomated2024}, the authors propose a framework that, starting from a conceptual model designed according to a meta-model, uses transformation rules to determine the most appropriate NoSQL model for implementation.  Similar to our approach, both data and schema are transformed. However, unlike our method, their goal is to adapt a general model to a specific NoSQL or relational model.  It would be worth exploring these ideas as a post-processing to our approach: from our target grammar $G_T$ and its associated instance $I_T$, find the most appropriate database model (e.g. many relationships might suggest a graph model).

To take our method a step further and also produce a user-friendly final schema, we might consider how to
\begin{enumerate*}
\item assign semantic names to our structures for better clarity (as in~\cite{barretAbstraGenericAbstractions2022})
  and

\item explore (as in~\cite{maliFACTDMFrameworkAutomated2024}) which database model best suits our schema.
  But this is out of the scope of this paper.
\end{enumerate*}

%

\section{Background}
\label{sec:Back}

This section reviews ordered trees and formal grammars, highlights the role of trees in representing linguistic structures, and considers tree rewriting rules for transformation and editing.
  
\begin{definition}[Ordered tree]
    \label{def:struct:tree}
     A tree $T = (D, l)$ consists of a domain  $D$ and a labelling function $l$.
     The domain $D$  is a subset of $(\mathbb{N})^*$ ( i.e.  a set of integer sequences of the form $x.y.z$). 
     The labelling function  $l : D \to \Sigma \cup \{\lambda\}$ maps elements of $D$ to labels from a set $\Sigma$ or  a special symbol $\lambda$.
     The domain $D$ satisfies the following properties:
     \begin{enumerate*}
        \item $D$ is closed under prefixes, i.e. for $u, u' \in (\mathbb{N})^*$ if $u$ is a prefix of $u'$ and $u' \in D$, then $u \in D$, and
        \item For all $u \in \mathbb{N}^*$ and $j \in \mathbb{N}$ if $u.j \in D$ then for all $i \in \mathbb{N}$ such that $0 \leq i < j$ we have $u.i \in D$.
     \end{enumerate*}

    Each element of $D$ is called \emph{position}.
    For a node $n$ at position $p$, $|p|$ defines the length of the sequence, also called the \emph{depth} of $n$.
    The root of a tree is at position $\epsilon$ and is labeled with the special symbol $\lambda$, i.e.  $l(\epsilon) = \lambda$.
    An empty tree is therefore defined by $T = (\{\epsilon\}, \langle \epsilon \mapsto \lambda \rangle)$.
    We write $v \prec u$ if $u = v.i$ for some $i \in \mathbb{N}$ where the node at position $v$ is the \emph{parent} of the node at $u$, and $u$ is the \emph{child} of $v$.
    We write $v \prec^* u$ if $\exists v'$ such that $u = v.v'$, meaning that $v$ is a \emph{prefix} of $u$.
    A node $n$ at position $u$ is a \emph{descendant} of a node $m$ at position $v$ if and only if $v$ is a direct prefix of $u$ (denoted $v \preceq u$) or indirect (denoted $v \preceq^* u$), conversely $m$ is an \emph{ancestor} of $n$.
    A \textit{leaf} is a node at a position $u$ such that $u.0 \notin D$, i.e.  a node with no children.
    We note $\mathbb{T}$ the set of all trees.
    \qed
\end{definition}

\begin{definition}[Sub-tree]
    Given a tree $T = (D, l)$, a sub-tree of $T$ at position $u \in D$ is denoted by $T|_u = (D', l')$ and has the following properties:
    \begin{enumerate*}
        \item $D' \subseteq D$ such that $\forall v \in D' ~ u \preceq^* v$ and
        \item $l' = \langle v \mapsto l(v) \mid v \in D' \rangle $.
    \end{enumerate*}
    Moreover, if $t = T|_u$ is a sub-tree, we denote by $t' = P_i^t$ the $i$th 
    tree-ancestor of $t$ when $t' = T|_v$, $u = vw$ and $|w| = i$.
    We note $\mathbb{ST}$ the set of all sub-trees.\qed
\end{definition}


\begin{example}
    Let $T = (D, l)$  be a tree with  $D=\{\epsilon, 0, 1, 1.0, 1.1\}$.
    The node labels are defined as follows: 
 $l(\epsilon) = root$,  $l(0)=child1$ (the left child of the root); $l(1)=child2$   (the right child of the root), and
for the  children of node at position $1$, we have $l(1.0)=grandchildren1$  and  $l(1.1)=grandchildren$.
       $T|_{1} = (D', l')$ is  a sub-tree of $T$.
    Note that  $T|_{1}$ is not a tree because $D'= \{1, 1.0, 1.1\}$ does not respect the conditions of Definition ~\ref{def:struct:tree}.
    Here, $P_1^{T|_{1}}$ coincides with $T$.\qed
\end{example}

\begin{definition}
    \label{def:struct:pre:cfg}
    A context-free grammar (CFG)  $G= (N, T, R, S)$ is a quadruplet where $N$ is a finite set of non-terminal symbols; $T$ is a finite set of terminal symbols; $R$ is a finite set of production rules, and $S \in N$ is the initial symbol.
    A production rule is defined by the form $X \to \alpha$, where $X \in N$ and $\alpha$ is a string of terminal and non-terminal symbols.
    As syntactic sugar, a condensed CFG,  allows production rules of the form $X \to \alpha^+$, where $\alpha^+$ is a regular expression that repeats the string $\alpha$ one or more times.
    This is equivalent to the rules $X \to \alpha$ and $X \to \alpha~X$.\qed
\end{definition}

A parse tree, also known as a derivation tree or syntax tree, describes how the starting symbol of a grammar $G$ derives a word in the language.
When a non-terminal $U$ is associated with a production rule $U \to X ~ Y ~ Z$, the derivation tree will contain an internal node labelled $U$ with three children, $X$, $Y$ and $Z$, arranged from left to right.
Each internal node represents a non-terminal symbol of the $G$ grammar, while the leaves represent the terminal symbols of $G$.
The links between the nodes illustrate how the symbols are derived from each other.
The derivation tree is built recursively, following the production rules of the grammar. It starts with a root node corresponding to the initial symbol of the grammar, and at each level of the tree the nodes are replaced by symbols according to the production rules.

\begin{definition}[Parse Tree]
    Given a grammar $G = (N, T, R, S)$, a parse (or derivation) tree of $G$ is a tree which satisfies the following properties:
    \begin{enumerate}[leftmargin=*]
        \item The root is designated by the start symbol $S$, i.e. $l(\epsilon) = S$ ;
        \item Each leaf $f$ is denoted by a terminal symbol, i.e. $l(f) \in T$ ;
        \item Each internal node $x$ is denoted by a non-terminal symbol, i.e. $l(x) \in N$ ;
        \item If $U$ is a non-terminal used as a label of an internal node $x$ and $X_1, \dots, X_n$ are the labels of the children of $n$ from left to right, then there exists a production rule $U \to X_1 ~ \dots ~ X_n$ in $R$.
        The labels $X_1, \dots, X_n$ represent a sequence of terminal and non-terminal symbols.\qed
    \end{enumerate}
\end{definition}


\begin{example}
  \label{ex:struct:cfg}
  Let $G = (\{P\}, \{0,1\}, R, P)$ where $R$ is the set of production rules
  $\{P  \to 0 \mid 1 \mid  P~0  \mid   P~1\}$.  The grammar produces binary numbers.
  For instance, the parse tree for $0011$ is
  \footnotesize
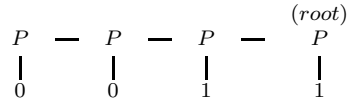
  \[
    \begin{array}{ccccccc}
      &&&&&& (root)\\
      P  &\horzbar&  P & \horzbar&   P & \horzbar &   P\\
      \vertbar && \vertbar &  & \vertbar    & & \vertbar \\
      0    && 0  & & 1  & & 1
    \end{array}
  \]
  \normalsize
  \qed
\end{example}

An attribute grammar extends a CFG by adding semantic information, stored in attributes tied to the grammar's terminal and non-terminal symbols. Attribute values are computed through rules linked to the grammar's productions. For each non-terminal in a CFG $G$, there are two sets of attributes:
\begin{enumerate*}
    \item \textit{synthesized attributes}, which pass information from the leaves to the root of a derivation tree and
    \item \textit{inherited attributes}, which pass information from the root to the leaves.
\end{enumerate*}
Each production is paired with semantic rules that define how to calculate the output attribute -- synthesized attributes of the left-hand non-terminal and inherited attributes of the right-hand non-terminals -- based on the input attributes. The following definition formalizes this concept.

\begin{definition}[Attribute Grammar]
\label{def:struct:G-attr}
An attribute grammar~\cite{knuthSemanticsContextfreeLanguages1968} is a CFG $G = (N, T, R, S)$ with a set of semantic rules $\Phi_r$ added to each production rule $r \in R$.
Each symbol $X \in (N \cup T)$ is associated to a finite set of attributes $A(X)$ consisting of two disjoint subsets of attributes:
\begin{enumerate*}[label=(\roman*)]
    \item $A_\uparrow(X)$ for synthesized attributes where $\forall X \in T , ~ A_\uparrow(X) = \emptyset$, and
    \item $A_\downarrow(X)$ for inherited attributes where, for the initial symbol $S$, $A_\downarrow(S) = \emptyset$
\end{enumerate*}

Each attribute $a \in A(X)$ has a (potentially infinite) set of possible values $V_a$ from which a value is selected for each occurrence of $X$ in the derivation tree.
The production rule $r$ has the form $X_0 \to X_1, \dots, X_n$ where $n \ge 1$, $X_0 \in N$ and $X_i \in (N \cup T)$ for $1 \le i \le n$.
A semantic rule $\varphi \in \Phi_r$  associated with $r$ is a function such that  $a=\varphi (b_1, \dots, b_k)$ for each output attribute $a$ of $r$ where
$b_i$ ($1 \leq k$) are input attributes of $r$. This rule computes the value of an attribute of $X_j$ from the attributes of the symbols $X_0, \dots, X_n$.
If $a$ is an attribute of $X_0$, it is a synthesized attribute.
If $a$ is an attribute of $X_j$ ($1 \leq j \leq n$), it is an inherited attribute.\qed
\end{definition}

\begin{example}
\label{ex:struct:cfg-attr}
Let  $G'$  be an  attribute grammar  built from $G$ of Example~\ref{ex:struct:cfg}
by associating  a synthesized attribute $val$ with the non-terminal $P$ and providing rules to compute the value of  $val$ relative to the value of the previously computed attribute $val'$ associated with the right side of the production rule.

\noindent
\begin{minipage}[c]{.3\linewidth}
    \begin{align*}
        P_{val} & \to 0 & [val \gets 0] \\
        P_{val} & \to 1 & [val \gets 1]
    \end{align*}
\end{minipage}%
\quad\vline%
\begin{minipage}[c]{.65\linewidth}
    \begin{align*}
        P_{val} & \to P_{val'}~0 & [val \gets 2 * val'] \\
        P_{val} & \to P_{val'}~1 & [val \gets 2 * val' + 1]
    \end{align*}
\end{minipage}\\

In the production rules, the attributes are indicated as subscripts and the semantic rules are presented in square brackets to the right of the production rule.
The grammar $G'$ associates the decimal value with a binary number.
The parse tree for $0011$
\footnotesize
\[
  \begin{array}{ccccccc}
    &&&&&& (root)\\
    P(val=0)  &\horzbar&  P(val=0) & \horzbar&   P (val=1) & \horzbar &   P (val=3)\\
    \vertbar && \vertbar &  & \vertbar    & & \vertbar \\
    0    && 0  & & 1  & & 1
  \end{array}
\]
\normalsize
shows, for each level, the computed value of the attribute $val$.\qed


\end{example}

\emph{S-attribute} grammars are attribute grammars containing only synthesized attributes, making them simpler and easier to verify through bottom-up propagation. 

\begin{definition}[Meta-grammar]
A meta-grammar $\mathbb{G} = (N, T, R, $ $S)$ is an S-attribute grammar where $N$ is the set of meta-non-terminals, $T$ is the set of meta-terminals, $R$ is the set of production rules, and $S \in N$ is the start symbol.
$\mathbb{G}$ specifies the syntax for production rules of condensed CFGs. Words in the language recognized by $\mathbb{G}$ are lists of production rules for condensed CFGs. 

A synthesized attribute $\gamma$ is used to verify that each derivation produces a valid condensed CFG. If $S_\gamma = \top$, the derivation is valid; if $S_\gamma = \bot$, it is invalid.

Semantic rules in production rules $r \in R$ fall into two types:
\begin{itemize}
\item $a \gets \alpha$, where $a$ is an attribute and $\alpha$ is a formula on attributes in $r$;
\item $\gamma \gets \beta$, where $\beta$ is a logical formula on attributes in $r$.
\end{itemize}
For clarity, we will omit the part $\gamma \gets$ and rules of the form $a \gets a$ in this paper.
\qed
\end{definition}

\vspace{0.2cm}
Syntax trees (as the on in Figure~\ref{fig-trees-ex1-stx}) are often used to represent texts, serving as derivation trees for the grammar of the natural language. In linguistics, they depict the syntactic structure of a sentence, highlighting the hierarchical relationships between words or word groups and their roles within parts of speech (PoS) such as nouns (NN), verbs (VBD), adjectives (ADJ), and determiners (DT). The leaves of the tree represent the lexical units (words), while the intermediate nodes correspond to abstract structures like verb phrases (VP) or noun phrases (NP).

Tree rewriting, or term rewriting, involves transforming trees into other trees using specific rewriting rules. In our context, these rules allow the formalization of tree transformations, with the goal of organizing information in a more structured manner while disregarding unnecessary elements.
In the following we recall the basis of tree rewriting.

\begin{definition}[Hedge]
    A hedge is a possibly empty sequence of trees, represented as $h = [t_0, \dots, t_n]$, with $|h|$ indicating the number of trees (i.e. $|h| = n + 1$).
    A substitution, denoted $\sigma$, is a bijective mapping from a set of variables $V$ to a set of hedges and from a set of labels to a set of sub-trees, homomorphically extended to trees.\qed
\end{definition}

\begin{definition}[Rewriting rule]
    A rewriting rule on a tree specifies how a tree $t$ can be rewritten as $t'$ at a given position $u$. It consists of a left-hand side (LHS), representing a pattern, and a right-hand side (RHS), representing the transformation, written as $LHS \to RHS$. The pattern is a subtree formed from the set $\Sigma \cup V \cup \{\lambda\}$, where $\Sigma$ is the set of labels of $t$, $V$ is a set of variables, and $\lambda$ is the root symbol. A morphism maps variables from LHS to RHS, enabling the transformation.

    The rule applies by substituting $\sigma$, a subtree of $t$ at position $u$, into the LHS pattern. This creates a correspondence between elements of the LHS and a subtree of $t$. The application of the rule is denoted as $t \mapsto_{[u,LHS \to RHS, \sigma]} t'$, where $t|_u = \sigma(LHS)$ and $t'|_u = \sigma(RHS)$.

    A rewriting rule may also include application conditions that specify when the rule can be applied, such as constraints on node or edge attributes or topological requirements.\qed
\end{definition}

\begin{figure}[htb]
    \centering
    \begin{adjustbox}{max width=\linewidth}
        \begin{forest}
            for tree={s sep=1em,l sep=.5em,l=.5em,fit=tight},
            where n children=0{tier=word}{}
            [$\lambda$ [$X$,baseline [$C$] [$B$] [$A$]] [$Y$ [$D$]]]
        \end{forest}
        \hspace{2em}
        {\huge$\Longrightarrow$}
        \hspace{2em}
        \begin{forest}
            for tree={s sep=1em,l sep=.3em,l=.5em,fit=tight},
            where n children=0{tier=word}{}
            [$\lambda$ [$X$,baseline [$C$] [$B$]] [$Y$ [$D$]]]
        \end{forest}
    \end{adjustbox}
    \caption{Example of the application of $rule(0.2)$ on a tree}
    \label{fig:sch:pre:rewritting:ex}
\end{figure}
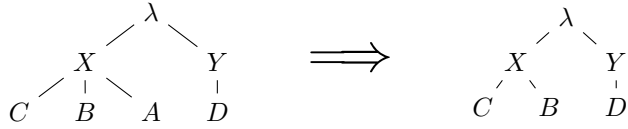

\begin{example}
    Let $\{X, Y, A, B, C, D\} \subseteq \Sigma$ be a set of labels (non-terminals) in trees. 
    Consider the rewriting rule $rule(u.i)$ with the constraint $|\sigma(x)| = i$ :
    \raisebox{-.4\height}{\small\begin{forest}
        for tree={s sep=.3em,l sep=.2em,l=.5em,fit=tight}
        [$U$ [$x$] [$A$]]
    \end{forest}}
    $\rightarrow$
    \raisebox{-.4\height}{\small\begin{forest}
        for tree={s sep=.3em,l sep=.3em,l=.2em,fit=tight}
        [$U$ [$x$]]
    \end{forest}}

    Here, $U$ is the node at position $u$ in the target tree, and $x \in V$ is a variable. 
    The rule applies to a tree $T$ if there is a sub-tree $t = T|_u$ with a substitution $\sigma$ such that $\sigma(x) = [t|_0, \dots, t|_{i-1}]$ and $l(u.i) = A$.
    Applying the rule deletes the sub-tree labeled $A$ from $T$.
    Figure~\ref{fig:sch:pre:rewritting:ex} show the application of $rule(0.2)$ with $u=0$ and $i=2$ where $\sigma(x) = [C, B]$.
    \qed
\end{example}

\section{Automatic Structuring}
\label{sec:autoStruc}

Algorithm~\ref{algo:struct:Main}  summarizes our process for structuring textual data.
Textual data is represented by trees, from which grammars are extracted to serve as a general representation. The structure that emerges from this iterative process is a grammar that plays the role of a database schema, together with its corresponding parse tree, which corresponds to a database instance. The resulting grammar can be mapped to various database models.

As input, Algorithm~\ref{algo:struct:Main} receives:
\begin{enumerate*}
    \item an instance $I_0$ corresponding to a forest of syntax trees generated from text sources, which have been merged into a single tree with a common root;
    \item the meta-grammar  $\mathbb{G}$ and
    \item a set of entities $\textbf{E}$.
\end{enumerate*}

We define a \textit{named entity} as  a tuple $E = (entityName,$ $startToken,$ $endToken)$ where $entityName$ is the name or type of the named entity, $startToken$ is the index of the token that marks the beginning of the entity, and $endToken$ is the index of the token that marks the end of the entity.
$L_{tokens}(E)$ corresponds to the sequence of tokens forming part of the entity, defined as $L_{tokens}(E) = [startToken, \dots, $ $endToken]$.
Named entities represent real-world objects and are instances of a class, with "Paris" as an example of a "City" entity. 
Algorithm~\ref{algo:struct:Main}  outputs a target grammar $G$ valid with respect to $\mathbb{G}$.

%
%

The iterative process can be summarized as follows.
It begins with an enrichment step (line~\ref{Main-ligne:enrichsimplify}) where entities and relationships are added as internal nodes to the syntax trees, followed by the removal of redundancies to simplify the structure. Next, the grammar is extracted (line~\ref{Main-ligne:extract}) and checked against a pre-established meta-grammar  (line~\ref{Main-ligne:while}). If the resulting grammar is not valid, tree transformations are applied. This involves computing equivalence classes for non-terminals (line~\ref{Main-ligne:computeeqclasses}) and then unifying and structuring equivalent sub-trees according to the meta-grammar $\mathbb{G}$ (line~\ref{Main-ligne:rewrite}). 
Then, a new grammar is extracted from the new instance (line~\ref{Main-ligne:extract2}) and the while loop proceeds with verification.

\begin{algorithm}[htb]
    \caption{AlgoStructMain($I_0$ , $\mathbb{G}$, $\textbf{E}$)}
    \label{algo:struct:Main}
    
    $I \gets EnrichSimplify(I_0, E)$ \label{Main-ligne:enrichsimplify} \\
    $G \gets ExtractGrammar(I)$\label{Main-ligne:extract} \\
    
    \While{$G$ not valid wrt $\mathbb{G}$  \label{Main-ligne:while}}{
     $ComputeEqClasses(I)$  \label{Main-ligne:computeeqclasses} \\
             $I \gets Rewrite(I)$ \label{Main-ligne:rewrite}\\
             $G \gets ExtractGrammar(I)$\label{Main-ligne:extract2}
     }   
       
    \Return{$G$}
\end{algorithm}

In the following sections, we outline each step of our approach. First, we define the meta-grammar $\mathbb{G}$, followed by an explanation of each step of Algorithm~\ref{algo:struct:Main}. 

\subsection{Meta-Grammar: generic database schema definition}
\label{sec:MG}
The purpose of the meta-grammar is to define the core concepts of a database model, in a generic manner (Section~\ref{sec:Over}). These concepts -- attribute (or entities), group, relation, and collection -- must now be expressed   by trees to enable their identification within the tree structure representing our data instance $I$.



Table~\ref{table:struct} presents our meta-grammar $\mathbb{G}$, an attribute grammar that defines valid data structures. Meta-non-terminals are indicated by angle brackets $\langle \cdot \rangle$, while the semantic rules are shown on the right side of the table within square brackets $[ \cdot ]$.

The first production meta-rule of $\mathbb{G}$ (\ref{meta:start}) indicates that the target grammar $G$ is defined by an initial rule,  followed by a possibly empty list of rules.
The initial rule generated by $\mathbb{G}$ (meta-rule~\ref{meta:root}) contains the symbol $\lambda$ (initial non-terminal of $G$) on its left-hand side.
Its right-hand side is defined by the meta-rules \ref{meta:rootList:1}-\ref{meta:rootList:6} which specify the construction of a series of $G$ non-terminals.
These non-terminals are: \emph{ENT}, \emph{GROUP}, \emph{REL} and \emph{COLL}, representing, respectively, entities, groups of entities, relations between groups and collections of groups or relations.
To distinguish each structure specific to a $G$ grammar, we associate a $name$ attribute with each non-terminal.
The attributes of $\mathbb{G}$ are synthesized and represent lists of names ($name$) used as identifiers:
\begin{itemize}
  \item $eL$ (or $eL'$): list of entity names;
  \item $gL$ (or $gL'$) : list of group names;
  \item $rL$ (or $rL'$) : list of relation names;
  \item $cgL$ (or $cgL'$) : list of group collection names;
  \item $crL$ (or $crL'$) : list of relation collection names.
\end{itemize}

They are initialized in a bottom-up fashion.
Syntax rules are used to check that a name is unique.
For example, if the meta rules \ref{meta:entList:1}-\ref{meta:entList:2} are applied, the list $eL$ gets entity names that are unique.
This is also the case for all other lists: the uniqueness of the name of a new non-terminal is guaranteed by the semantic rules.
It is also important to note that these semantic rules ensure that any non-terminal appearing to the right of a production rule in the $G$ grammar must have a rule defining it.
For example, if the meta rule \ref{meta:start} is applied, all elements in the list $gL'$ must be present in $gL$.

\begin{landscape}
  \begin{table}[h]
    \centering
    \begin{adjustbox}{max width=\linewidth,max height=\textheight,valign=c}
      \begin{minipage}{1.1\linewidth}
        \small
        \begin{align}
          \epsilon                                    & ::= \langle root_{eL',gL',cgL',rL',crL'} \rangle ~\textsc{eol}~ \langle ruleList_{eL,gL,cgL,rL,crL} \rangle  & [eL' \subseteq eL; gL' \subseteq gL; cgL' \subseteq cgL; rL' \subseteq rL; crL' \subseteq crL]         \label{meta:start}      \\
          \langle root_{eL,gL,cgL,rL,crL} \rangle     & ::= \lambda \to \langle rootList_{eL,gL,cgL,rL,crL} \rangle                                                                                                                                                           \label{meta:root}       \\[1em]
          \langle rootList_{eL,gL,cgL,rL,crL} \rangle & ::= \epsilon                                                                                                 & [eL \gets \emptyset; gL \gets \emptyset; cgL \gets \emptyset; rL \gets \emptyset; crL \gets \emptyset] \label{meta:rootList:1} \\
          & ~~ \mid ~ ENT_{name}  ~ \langle rootList_{eL',gL,cgL,rL,crL} \rangle                                         & [name \notin eL'; eL \gets \{name\} \cup eL']                                                          \label{meta:rootList:2} \\
          & ~~ \mid ~ GROUP_{name} ~ \langle rootList_{eL,gL',cgL,rL,crL} \rangle                                        & [name \notin gL'; gL \gets \{name\} \cup gL']                                                          \label{meta:rootList:3} \\
          & ~~ \mid ~ REL_{name}  ~ \langle rootList_{eL,gL,cgL,rL',crL} \rangle                                         & [name \notin rL'; rL \gets \{name\} \cup rL']                                                          \label{meta:rootList:4} \\
          & ~~ \mid ~ COLL_{name} ~ \langle rootList_{eL,gL,cgL',rL,crL} \rangle                                         & [name \notin cgL'; cgL \gets \{name\} \cup cgL']                                                       \label{meta:rootList:5} \\
          & ~~ \mid ~ COLL_{name} ~ \langle rootList_{eL,gL,cgL,rL,crL'} \rangle                                         & [name \notin crL'; crL \gets \{name\} \cup crL']                                                       \label{meta:rootList:6} \\[1em]
          \langle ruleList_{eL,gL,cgL,rL,crL} \rangle & ::= \epsilon                                                                                                 & [eL \gets \emptyset; gL \gets \emptyset; cgL \gets \emptyset; rL \gets \emptyset; crL \gets \emptyset] \label{meta:ruleList:1} \\
          & ~~ \mid ~ \langle entity_{name}          \rangle ~\textsc{eol}~ \langle ruleList_{eL',gL,cgL,rL,crL} \rangle & [name \notin eL'; eL \gets \{name\} \cup eL']                                                          \label{meta:ruleList:2} \\
          & ~~ \mid ~ \langle group_{name, eL'}      \rangle ~\textsc{eol}~ \langle ruleList_{eL,gL',cgL,rL,crL} \rangle & [name \notin gL' \land eL' \subseteq eL; gL \gets \{name\} \cup gL']                                   \label{meta:ruleList:3} \\
          & ~~ \mid ~ \langle relation_{name, gL'}   \rangle ~\textsc{eol}~ \langle ruleList_{eL,gL,cgL,rL',crL} \rangle & [name \notin rL' \land gL' \subseteq gL; rL \gets \{name\} \cup rL']                                   \label{meta:ruleList:4} \\
          & ~~ \mid ~ \langle collGrp_{name,grpName} \rangle ~\textsc{eol}~ \langle ruleList_{eL,gL,cgL',rL,crL} \rangle & [name \notin cgL' \land grpName \in gL; cgL \gets \{name\} \cup cgL']                                  \label{meta:ruleList:5} \\
          & ~~ \mid ~ \langle collRel_{name,relName} \rangle ~\textsc{eol}~ \langle ruleList_{eL,gL,cgL,rL,crL'} \rangle & [name \notin crL' \land relName \in rL; crL \gets \{name\} \cup crL']                                  \label{meta:ruleList:6} \\[1em]
          \langle group_{name, eL} \rangle            & ::= GROUP_{name} \to \langle entList_{eL} \rangle                                                                                                                                                                     \label{meta:group}      \\
          \langle collGrp_{name,grpName} \rangle      & ::= COLL_{name} \to GROUP_{grpName}^+                                                                                                                                                                                 \label{meta:collGroup}  \\[1em]
          \langle relation_{name, gL} \rangle         & ::= REL_{name} \to GROUP_{name1} ~ GROUP_{name2}                                                             & [name1 \neq name2; gL \gets \{name1, name2\}]                                                          \label{meta:rel}        \\
          \langle collRel_{name,relName} \rangle      & ::= COLL_{name} \to REL_{relName}^+                                                                                                                                                                                   \label{meta:collRel}    \\[1em]
          \langle entList_{eL} \rangle                & ::= ENT_{name}                                                                                               & [eL \gets \{name\}]                                                                                    \label{meta:entList:1}  \\
          & ~~ \mid ~ ENT_{name} ~ \langle entList_{eL'} \rangle                                                         & [name \notin eL'; eL \gets \{name\} \cup eL']                                                          \label{meta:entList:2}  \\
          \langle entity_{name} \rangle               & ::= ENT_{name} \to \langle data \rangle \label{meta:entity}
        \end{align}
      \end{minipage}
    \end{adjustbox}
    \caption{Meta-grammar $\mathbb{G}$ using BNF format}
    \label{table:struct}
  \end{table}
  \end{landscape}



\begin{figure}[htb]
  \centering
  \begin{adjustbox}{max width=\linewidth}
    \begin{forest}
      for tree={s sep=.3em,l sep=.5em,l=1em,fit=tight,align=center},
      [$\epsilon$
        [{$\langle root \rangle$\\$\color{blue}crL=\{1\}$}
          [$\lambda$]
          [$\to$,before computing xy={s/.average={s}{siblings}}]
          [{$\langle rootList \rangle$\\$\color{blue}crL'=\{1\}$}
            [$COLL_1$]
            [$\langle rootList \rangle$
              [$\epsilon$]
            ]
          ]
        ]
        [\textsc{eol},before computing xy={s/.average={s}{siblings}}]
        [{$\langle ruleList \rangle$\\$\color{blue}crL=\{1\}$, $\color{blue}gL=\{1, 2\}$\\$\color{blue}rL=\{1\}$, $\color{blue}eL=\{1, 2, 3\}$}
          [{$\langle collRel \rangle$\\$\color{blue}name=1$\\$\color{blue}relName=1$}
            [$COLL_1$]
            [$\to$,before computing xy={s/.average={s}{siblings}}]
            [$REL_1^+$]
          ]
          [\textsc{eol},before computing xy={s/.average={s}{siblings}}]
          [\dots]
        ]
      ]
    \end{forest}
  \end{adjustbox}
  \caption{Extract of a derivation of $\mathbb{G}$}
  \label{fig:metaDerivation}
\end{figure}
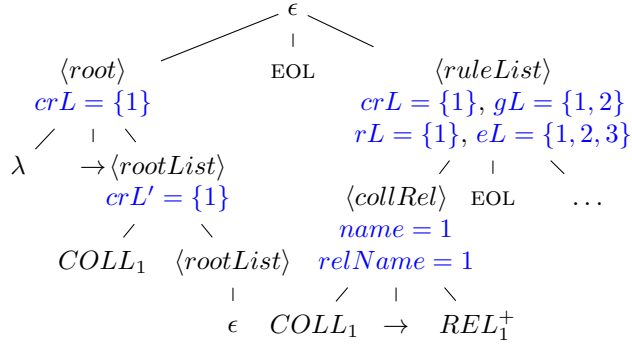

\begin{example}
  \label{ex:struct}
  Consider the grammar $G$ from figure~\ref{fig:metaConcepts} and the derivation of $\mathbb{G}$ that leads to the rule $\lambda \to COLL_1$.
  By applying meta-rule~\ref{meta:root}, we derive the rule $\lambda \rightarrow \langle rootList \rangle$, where the right-hand side contains a meta-non-terminal.
  Next, applying meta-rule~\ref{meta:rootList:6} results in the intermediate rule $\lambda \rightarrow COLL_1 ~ \langle rootList \rangle$, and finally, meta-rule~\ref{meta:rootList:1} produce $\lambda \rightarrow COLL_1$.
  The set of production rules for the grammar $G$ is defined by meta-rules~\ref{meta:ruleList:1}-\ref{meta:ruleList:6}, where each rule introduces a non-terminal for $G$.

  Figure~\ref{fig:metaDerivation} presents a partial derivation of $G$ from the meta-grammar $\mathbb{G}$.
  Attributes are displayed in blue, excluding $\gamma$ and empty-set attributes for clarity.
  Note that the semantic rules of $\mathbb{G}$ impose constraints to ensure  that every non-terminal in the target grammar $G$ is properly defined, a requirement for constructing a valid grammar.
  In this context, Figure~\ref{fig:metaDerivation} shows,  on the  the left-hand side of the root, that  $crL' = \{1\}$, signifying that $COLL_1$ is referenced in the root rule. On the right-hand side, $\langle ruleList \rangle$ holds $crL = \{1\}$.
  Since $crL' \subseteq crL$, this confirms that every non-terminal appearing in the root rule of $G$ has a corresponding defined production rule, ensuring a valid derivation.\qed
  
  
\end{example}

\subsection{Tree Enrichment and Simplifications}
\label{sec:EnrichTrees}
The enrichment step involves simplifying the syntax trees and adding semantic information, mainly formalised by tree rewriting rules. This section gives an overview of these operations.

\subsubsection{Simplifying Conjunction Sub-Trees}
Syntactic trees are generated using parsers, typically trained on treebanks.
However, due to linguistic variations, syntactic annotations across treebanks are not standardized.
In both French and English, coordinating conjunctions can be represented using either recursive or flat structures.
In this paper, we utilize the English and French parsers provided by CoreNLP, where the English parser tends to favor a flatter structure, placing coordinated elements at the same hierarchical level.
Since sub-trees for coordinating conjunctions are suitable candidates for \textit{collection} structures (mentioned in Section~\ref{sec:MG}), we rewrite the conjunctions sub-trees to this flatter representation to better align with our meta-model.

\subsubsection{Enriching Trees}
 To incorporate an entity in a syntax tree $T$ means to create an \emph{ENT}-labeled sub-tree, specifying the entity type (e.g., Person, Country, Disease) and containing leaves for the entity's tokens.
 
\begin{definition}[Ordered entity subtree]
  \label{def:struct:entity-tree}
  Let $T = (D, l)$ be a syntax tree, $E = (entityName,$ $startToken,$ $endToken)$ be an entity and $L_{tokens}(E)$ be the list of indices of $E$'s tokens.
  An ordered entity sub-tree is a tuple $E_T = (entityName,$ $L_{tree}(T, E))$ where
  $L_{tree}(T, E)$ is the sequence of positions of the tokens of $E$ in the tree $T$ such that 
  $L_{tree}(T, E) = [u.b,$ $\dots,$ $u.e]= [treePos(T, \linebreak startToken),$ $\dots,$  $treePos(T, endToken)]$ 
  with $u$ being the position common to all the tokens of $E$ and $treePos : \mathbb{T} \times \mathbb{N} \to D$ 
  being a function which associates for each index of a token its position in the tree.
  $|E|$ is the size of the entity (or number of tokens) such that :
  $|E| = |L_{tokens}(E)| = |L_{tree}(T, E)| = (endToken - startToken) + 1$.\qed
\end{definition}
    
\begin{figure*}[htb]
  \centering
  \begin{subfigure}{.55\linewidth}
    \centering
    \begin{adjustbox}{max width=\linewidth,max height=.2\textheight}
      \begin{forest}
        for tree={s sep=.2em,l sep=.5em,l=1em,fit=tight}
        [U
          [$x_0$]
          [ENT,baseline
            [$z_0$]
            [ENT$_0$ [$y_0$]]
            [$z_1$]
            [$\dots$]
            [ENT$_n$ [$y_n$]]
            [$z_{n+1}$]
          ]
          [$x_1$]
        ]
      \end{forest}
      {\huge$\longrightarrow$}
      \begin{forest}
        for tree={s sep=.2em,l sep=.5em,l=1em,fit=tight}
        [U
          [$x_0$]
          [ER
            [ENT,baseline [$z_0$] [$y_0$] [$z_1$] [$\dots$] [$y_n$] [$z_{n+1}$]]
            [EC
              [ENT$_0$ [$y_0$]]
              [$\dots$]
              [ENT$_n$ [$y_n$]]
            ]
          ]
          [$x_1$]
        ]
      \end{forest}
    \end{adjustbox}
    \caption{$\textsf{unnest\_ent}(T, u.i)$ where $|\sigma(x_0)| = i$}
    \label{fig:sch:op:unnestEnt}
  \end{subfigure}
  \hfill
  \begin{subfigure}{.4\linewidth}
    \centering
    \begin{adjustbox}{max width=\linewidth,max height=.2\textheight}
      \begin{forest}
        for tree={s sep=.2em,l sep=.5em,l=1em,fit=tight}
        [U [$x$,baseline] [$y_1$ [$y_2$]] [$z$]]
      \end{forest}
      \hspace{1em}
      {\large$\longrightarrow$}
      \hspace{1em}
      \begin{forest}
        for tree={s sep=.2em,l sep=.5em,l=1em,fit=tight}
        [U [$x$,baseline] [$y_2$] [$z$]]
      \end{forest}
    \end{adjustbox}
    \caption{$\textsf{reduce}(T, u.i, S_{labels})$ where $|\sigma(x)| = i$ and $|\sigma(y_1)| = 1$ with $|\sigma(y_2)| = 1$ if $S_{label} = \emptyset$ ; $t(u.i) \notin S_{labels}$ if $S_{label} \neq \emptyset$}
    \label{fig:sch:op:reduce}
  \end{subfigure}
  \caption{Basic operations}
\end{figure*}
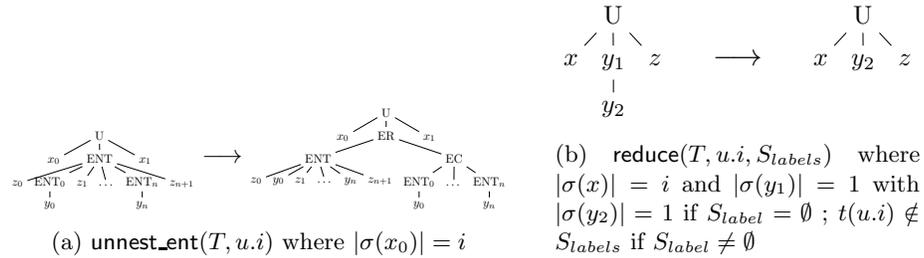

Entities correspond to the concept of attributes (Section~\ref{sec:Over}) in our generic database model, defined by the meta-grammar $\mathbb{G}$.
We maintain the classical notion of attributes; therefore, nested entities are considered invalid under $\mathbb{G}$.
To comply, nesting is represented as a relationship between the encompassing entity and the contained entities, forming a tree with two children.
This transformation is performed by the application of the tree rewriting rule 
\textsf{unnest\_ent} depicted in Figure~\ref{fig:sch:op:unnestEnt}.

\subsubsection{Simplifications}
The trees are simplified in two steps:
\begin{enumerate}[leftmargin=*]
  \item Sub-trees without entities are deleted.
  They are identified by checking the parents of all leaves in $T$.
  If they aren't labelled with an entity name, they are removed (e.g., sub-trees $T|_{0.0}$ and $T|_{1.0}$ in Figure~\ref{fig-trees-ex1-ent}).
  
  \item Nodes not labelled as entities and with only one child are deleted (e.g., nodes at position $0$, $1.0.0$ and $1.0.1$ in Figure~\ref{fig-trees-ex1-simp}).
  This is done using the rewriting rule \textsf{reduce} from Figure~\ref{fig:sch:op:reduce}.
\end{enumerate}
Figure~\ref{fig-trees-ex1-reduce} shows an enriched tree after  simplifications.

\subsection{Grammar Extraction}
\label{sec:extractG}


In lines~\ref{Main-ligne:extract} and~\ref{Main-ligne:extract2}  of Algorithm~\ref{algo:struct:Main}, we extract the grammar $G$ from a given instance $I$ and then verify its correctness against the meta-grammar $\mathbb{G}$ (line~\ref{Main-ligne:while}). The grammar $G$ is derived by calculating the quotient tree $S$, a hierarchical representation of $I$, used to obtain~$G$.

We recall that a partition of a set $X$ is a division of its elements into non-empty, disjoint subsets. An \textit{equivalence relation} on a set defines a partition, and every partition corresponds to an equivalence relation.
A set family $F$ is a partition of a set $X$ if and only if all the following conditions are satisfied:
$(i)$ $\emptyset \not\in F$ ;
$(ii)$  $\bigcup_{A \in F}A = X$ and
$(iii)$ $\forall A, B \in F~ (A \neq B) \Rightarrow ( A \cap B = \emptyset)$.
The sets of $F$ are called \textit{blocks}.
In graph theory, a quotient graph $Q$ of a graph $G$ is a graph whose vertices are blocks of a partition of the vertices of $G$ and where a block $A$ is adjacent to a block $B$ if at least one vertex of $A$ is adjacent to a vertex of $B$ with respect to the set of edges of $G$.
To construct the grammar from a tree, we introduce the definition of \textit{quotient tree}, which corresponds to a quotient graph with no cycles and no vertices with multiple parents.

\begin{figure}[htb]
    \centering
    \begin{subfigure}{.45\linewidth}
        \centering
        \begin{adjustbox}{max width=\linewidth}
            \begin{forest}
                for tree={s sep=.8em,l sep=.5em,l=1em,fit=tight},
                where n children=0{tier=word}{}
                    [$\lambda$
                        [$X$ [$a$] [$b$]]
                            [$X$,before computing xy={s/.average={s}{siblings}} [$b$] [$c$]]
                            [$Y$ [$a$]]
                    ]
            \end{forest}
        \end{adjustbox}
        \caption{Instance tree $T$}
        \label{fig:struct:quotient:ex:tree}
    \end{subfigure}
    \hfill
    \begin{subfigure}{.45\linewidth}
        \centering
        \begin{adjustbox}{max width=\linewidth}
            \begin{forest}
                for tree={s sep=.8em,l sep=.5em,l=1em,fit=tight},
                where n children=0{tier=word}{}
                    [$\lambda$
                        [$X^+$ [$a$] [$b$] [$c$]]
                            [$Y$ [$a$]]
                    ]
            \end{forest}
        \end{adjustbox}
        \caption{Quotient tree $Q_T$}
        \label{fig:struct:quotient:ex:quotient}
    \end{subfigure}
    \caption{Example of quotient tree computation}
    \label{fig:struct:quotient:ex}
\end{figure}
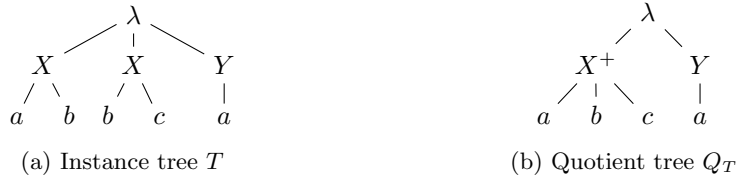

To extract the grammars $G_i$ we use the equivalence relation $R_l$ between the labels of a tree $T$, defined by $(\forall x, y \in D) ~ x ~ R_l ~ y \iff l(x) = l(y)$.
Intuitively, we obtain as equivalence classes a set of positions for each label present in the $T$ tree.
For example, in Figure~\ref{fig:struct:quotient:ex:tree}, $C_{\lambda} = \{\epsilon\}$ and $C_{X} = \{0,1\}$ are the equivalence classes for the labels $\lambda$ and $X$ respectively.

\vspace{0.2cm}
\noindent
To construct a quotient tree, we follow two steps:

\noindent
$\bullet$ \textit{ Compute the hierarchy of equivalence classes.} We define the function \textsf{Succ}, which, for a given class $C$, returns the set of equivalence classes containing at least one element that is a child of an element in $C$.

\begin{definition}[Function \textsf{Succ}]
    \label{def:struct:quotient:succ}
    Let $T = (D, l)$ be a tree and $R$ an equivalence relation.
    Let $D/R = \{C_0, \dots, C_n\}$ be the set of equivalence classes of $T$.
    Define the function \textsf{Succ} as
    $\textsf{Succ}(C) = \{C' \mid \exists u \in C, v \in C' \mbox{ such that } u \prec v \}$
    \qed
\end{definition}

\begin{example}
    \label{ex:struct:quotient:1}
    Let  $T = (D, l)$ be the tree Figure~\ref{fig:struct:quotient:ex:tree}.
    The first step in constructing the quotient tree $Q_T$ is to retrieve the equivalence classes of $D$ given by the relation $R_l$.
    We then obtain the following classes:  $C_\lambda  = \{\epsilon\}$, $C_X = \{0, 1\}$,  $ C_Y  = \{2\}$,
    $C_a  = \{00, 20\} $, $ C_b  = \{01, 10\}$, $ C_c  = \{11\} $.
    The construction of $Q_T$ involves recovering the hierarchy of equivalent sets.
    The algorithm starts with the class $C_\lambda$ and recursively traverses its successors.
    For each class, we have the following successors:
    $\textsf{Succ}(C_\lambda)  = \{C_X, C_Y\}$,
    $ \textsf{Succ}(C_X)  = \{C_a, C_b, C_c\} $,
    $ \textsf{Succ}(C_Y)  = \{C_a\} $,
    $\textsf{Succ}(C_a)  = \emptyset $,
    $ \textsf{Succ}(C_b)  = \emptyset $,
    $ \textsf{Succ}(C_c)  = \emptyset$. $\hfill\Box$
\end{example}

\noindent
$\bullet$ \textit{  Assign each equivalence class a position in the ordered tree.} 
The function \textsf{QDom} computes the domain of the quotient tree using the \textsf{Succ} function. \textsf{QDom} assigns each successor class of $C$ a relative position, $u.p$, where $u$ is the position of class $C$.
 In other words,
\begin{equation} \label{eq1}
    \small \textsf{QDom}(C, u)=\{(C_j, u.p) \mid C_j \in \textsf{Succ}(C) \mbox{ and } 0 \leq j \leq n-1\}
\end{equation}
where $n$ is the number of elements in $\textsf{Succ}(C)$.
It is worth noting that it is possible for an equivalence class to appear as the successor of more than one class.  In Example~\ref{ex:struct:quotient:1}, $C_a$ is a successor of both $C_Y$ and $C_X$, resulting in different positions for $C_a$.

To construct the quotient tree $Q_T = (Q_D, Q_l)$ from a tree $T$, we start by assigning the equivalence class $C_\lambda$ to the root of $Q_T$, i.e., $Q_D \gets \{\epsilon\}$ and $Q_l \gets \langle \epsilon \mapsto \lambda \rangle$. We also initialize a set \textit{classes} with pairs in the format \textit{(equivalence class, position in $Q_T$)}, denoted $(CL,u)$. While \textit{classes} is not empty, the algorithm iterates, selecting a pair $(CL,u)$, updating \textit{classes} with its successors computed using the \textsf{QDom} function, and updating the domain $Q_D$ with the new position $u$. The algorithm checks if the equivalence class $CL$ includes multiple positions with the same parent. 
Indeed, to construct a condensed CFG,
sibling nodes with the same non-terminal in a parse tree are compactly represented in the quotient tree with the ${}^+$ symbol. This involves checking for repeated non-terminals with the same parent and merging their derivations, allowing for trees with incomplete information, as shown in the following example.

\begin{example}\label{ex:struct:quotient:2}
    Let us consider the  result obtained in Example~\ref{ex:struct:quotient:1}.
    Function \textsf{QDom} is used to associate positions to equivalent classes.
    We start with $\textsf{QDom}(C_\lambda, \epsilon) = \{(C_X, 0), (C_Y, 1)\}$ and
    after successive applications of $\textsf{QDom}$ (see expression~(\ref{eq1})) we obtain:
    {\small\begin{align*}
        \textsf{QDom}(C_X, 0)  & = \{(C_a, 00), (C_b, 01), (C_c, 02)\} \\
        \textsf{QDom}(C_Y, 1)  & = \{(C_a, 10)\}                       \\
        \textsf{QDom}(C_a, 00) & = \emptyset \hspace{2cm}
        \textsf{QDom}(C_a, 10) = \emptyset                             \\
        \textsf{QDom}(C_b, 01) & = \emptyset \hspace{2cm}
        \textsf{QDom}(C_c, 02) = \emptyset
    \end{align*}}


Figure~\ref{fig:struct:quotient:ex:quotient} illustrates the obtained quotient tree.
    The node corresponding to class $C_a$ is duplicated in $Q_T$, because it is linked to two positions: $00$ and $10$.
     The node corresponding to class $C_X$ in $Q_T$ is marked with `${}^+$', indicating that $X$ can be repeated as a child of $\lambda$.
    This conclusion comes  from Example~\ref{ex:struct:quotient:1}:  $C_X = \{0, 1\}$ contains two positions with the parent at $\epsilon$.
  In the construction of $Q_T$,  after a first iteration, $classes= \{(C_X, 0), (C_Y,1)\}$.
    For   the pair $(C_X, 0)$, we have $successors = \{(C_a, 00), (C_b,01), (C_c,02)\}$.
    The class $C_X$ is not a singleton and its positions  have the same parent.\qed


\end{example}

A tree $T$ may represent an incomplete derivation of the condensed CFG $G_T$ obtained from a quotient tree $Q_T$.
For instance, if a production rule of $G_T$ is $X \to a~b~c$, the tree $T$ of Figure~\ref{fig:struct:quotient:ex:tree} is  accepted as a valid derivation only if  $c$ or $a$ are considered as missing values, which reflect omissions or errors in the syntactic analysis of natural language texts.
Incomplete information has been a challenge for database research (see, e.g, \cite{imielinskiIncompleteInformationRelational1984,Lib06,chabinConsistentUpdatingDatabases2020,chabinManagingLinkedNulls2023}).
The transformation of a quotient tree into a grammar is formally defined below.

\begin{definition}[Construction of a grammar from a quotient tree]
    The condensed CFG  $G_T$, obtained from the quotient tree $Q_T = (Q_D, Q_l)$ of $T$, is  defined  by the quadruplet $(N, T, P, \lambda)$ where:
    \begin{enumerate*}[label=(\roman*)]
        \item the set of non-terminals $N$, possibly decorated by ${}^+$, is the set of labels $Q_l(u)$ for any position $u \in Q_D$ which is not a leaf;
        \item the set of terminals $T$,   is the set of labels $Q_l(u)$ for any position $u \in Q_D$ which is a leaf;
        \item the set $P$ of production rules contains, for any position $u \in Q_D$ which is not a leaf, rules of the form $Q_l(u) \to Q_l(u.0), \dots Q_l(u.i)$ and
        \item $\lambda$ is the starting symbol.
    \end{enumerate*}
    \qed
\end{definition}

\begin{example}
    \label{ex:struct:quotient:3}
    From $Q_T$ in Example~\ref{ex:struct:quotient:2}, Figure~\ref{fig:struct:quotient:ex:quotient}, we obtain the grammar $G_T$ with the following rules:
    \begin{align*}
        \lambda & \to X^+ ~ Y & X & \to a ~ b ~ c & Y & \to a
    \end{align*}
\end{example}

\subsection{Computing Equivalence Classes}
\label{sec:EqClasses}
Algorithm~\ref{algo:struct:Main}, on line~\ref{Main-ligne:computeeqclasses}, modifies instance $I$ by identifying equivalence classes of sub-trees (non-terminals of the target grammar).
At this stage, trees have been enriched, simplified, or rewritten.
In the initial iteration, the tree contains named entity information, as shown in Figure~\ref{fig-trees-ex1-reduce}.

In our context, identifying equivalent sub-trees is essential for aggregating information.
The textual representation of a real-world object can take different forms, reflected in different entity sub-trees.
Besides, natural language often omits or implies information.
For example, the parse trees of \enquote{The patient takes 500 mg of Paracetamol} and \enquote{The patient takes Paracetamol every day} are different, but both represent a \textit{treatment}.

Determining sub-tree equivalence requires more than comparing entity labels, because natural language is ambiguous, and the same entity tree may represent different objects.
Context must also be taken into account.
We use the concept of regular equivalence from~\cite{whiteGraphSemigroupHomomorphisms1983}, where two vertices in a graph are equivalent if their neighbourhoods are equivalent.
For example, two people can be considered \emph{equivalent} (e.g. both representing a patient) if they are connected to \emph{equivalent} vertices, such as a disease or a treatment, even if those vertices are different.

To define the equivalence relation, we introduce a similarity measure between sub-trees.
A similarity measure is a symmetric function $f : \mathbb{ST} \times \mathbb{ST} \to [0,1]$ with $f(x, x) = 1$ for all $x \in \mathbb{ST}$.
Various measures  $f$ like Jaccard, Levenshtein, Jaro, or tree edit distance \cite{zhangSimpleFastAlgorithms1989} can be used.
The contextual similarity between two enriched sub-trees $x = T|_u$ and $y = T|_v$, denoted $sim_f(x, y)$, is computed as a weighted average of the recursive similarities provided by the function $f$ for each tree-ancestor.
The weights decrease as the distance from the tree-ancestor increases.
The formula for $sim_f(x, y)$ is given by the following equation, where $depth_{min}$ is the minimum depth of the sub-trees $x$ and $y$ (i.e. $depth_{min} = \min\{|u|, |v|\}$), and $P^x_i$ (or $P^y_i$) is the $i$-th tree-ancestor of $x$ (or $y$).
\begin{equation}
    sim_f(x, y) = \frac{\sum_{i=0}^{depth_{min}} \frac{1}{i + 1} \cdot f(P^x_i, P^y_i)}{\sum_{j=0}^{depth_{min}} \frac{1}{j + 1}} \label{eq:struct:sim}
\end{equation}

\begin{axiom}
    The function $sim_f$ is a weighted average of $f$.
    Therefore, $sim_f$ is symmetric, bounded by the interval $[0, 1]$ and for all $x \in \mathbb{ST}$, $sim_f(x, x) = 1$.\qed
\end{axiom}


\begin{figure}[htb] 
    \centering
    \begin{adjustbox}{max width=\linewidth}
        \begin{forest}
            for tree={s sep=1em,l sep=.5em,l=1em,fit=tight}
            [S
                [\dots]
                [CONJ,before computing xy={s/.average={s}{siblings}},where n children=0{tier=word}{}
                        [NP$_1$
                            [$X_1$
                                    [\textbf{ENT\_VALUE} [500]]
                                        [\textbf{ENT\_UNIT} [mg]]
                                ]
                                [\textbf{ENT\_DRUG} [Paracetamol]]
                        ]
                        [NP$_2$
                            [$X_2$
                                    [\textbf{ENT\_VALUE} [200]]
                                        [\textbf{ENT\_UNIT} [mg]]
                                ]
                                [\textbf{ENT\_FREQ} [every day]]
                        ]
                ]
                [\dots]
            ]
        \end{forest}
    \end{adjustbox}
    \caption{Extract of an enriched tree}
    \label{fig:struct:sim:ex}
\end{figure}
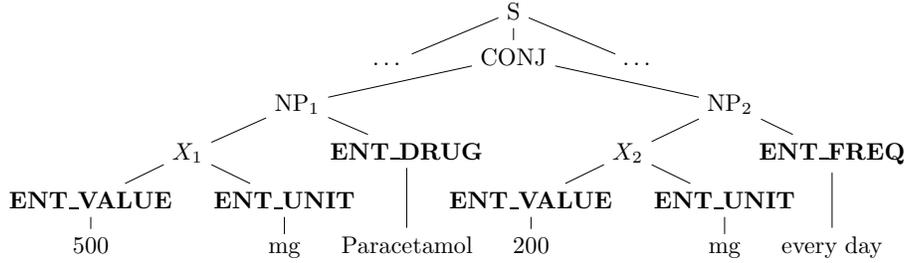

\begin{example}
    Let $T$ be the tree of Figure~\ref{fig:struct:sim:ex}.
    To define $f$, we use the Jaccard index ($J(X, Y) = \lvert X \cap Y \rvert / \lvert X \cup Y \rvert$) on the entity names present in the sub-tree.
    Although $f(\text{X}_1, \text{X}_2) = 1$, their contexts — one related to a drug (paracetamol) and the other to a frequency (every day) — suggest a similarity of less than $1$.
    The function $sim_f$ (equation~\ref{eq:struct:sim}) accounts for this difference.
    We find $f(\text{NP}_1, \text{NP}_2) = 0.75$ and continue recursively to the root, where the sub-trees are identical with a similarity of $1$.
    This results in :
    \begin{equation*}
        \small sim_f(\text{X}_1, \text{X}_2) = \frac{\overbrace{1 \times 1}^{\text{X}} + \overbrace{0.5 \times 0.75}^{\text{NP}} + \overbrace{0.33 \times 1}^{\text{CONJ}} + \overbrace{0.25 \times 1}^{\text{S}}}{1 + 0.5 + 0.33 + 0.25} \simeq 0.94
    \end{equation*}
    \qed
\end{example}

The example shows that although the two sub-trees are similar, they are not equivalent because they do not refer to the same objects.
Our similarity measure defines when two sub-trees are similar, based on a threshold set for each dataset.

\begin{definition}[Sub-tree similarity]
    \label{def:struct:sim}
%
    Given an enriched tree $T$, let $st_1 = T|_u$ and $st_2 = T|_v$  be two sub-trees. Let  $\tau \in [0, 1]$ be a threshold. We say that $st_1$ and $st_2$ are $\tau$-similar, denoted $st_1 \sim_\tau st_2$, if and only if $sim_f(st_1, st_2) \geq \tau$.\qed
\end{definition}

\begin{proposition}
    $\tau$-similarity is a \emph{reflexive} and \emph{symmetric} similarity relation.\qed
\end{proposition}

\begin{definition}[Sub-tree equivalence]
    \label{def:struct:equiv}
     Let $T$  be  an enriched tree. Given a  $\tau$-similarity relation  (Definition~\ref{def:struct:sim}),
    we define an equivalence relation between the sub-trees  
    $x = T|_u$ and $y = T|_v$ (denoted $x \equiv_\tau y$) by the following equation :
    \small
    \begin{equation}
        (\forall x, y \in \mathbb{ST}) ~ x \equiv_\tau y \iff x \sim_\tau y \lor (\exists z \in \mathbb{ST}) ~ x \equiv_\tau z \land y \equiv_\tau z
        \label{eq:struct:equiv}
    \end{equation}
    \normalsize
\end{definition}

\begin{proposition}
    The $\tau$-equivalence is an equivalence relation, that is, \emph{reflexive}, \emph{symmetric} and \emph{transitive}.
\end{proposition}

\begin{definition}[Equivalence classes]
    \label{def:eqClasses}
    Let $[x]_\tau$ denote the $\tau$-equivalence class of $x$, where $y \in [x]_\tau$  if and only if $y \equiv_\tau x$.
    For a tree $T = (D, l)$, $D/_{\equiv_\tau} = \{[x]_\tau \mid x \in D\}$ represents the quotient set (or partition) of $D$ by $\equiv_\tau$, i.e., it is the set of all $\tau$-equivalence classes of $D$.\qed
\end{definition}

Partitioning a set based on distance can be done using single-link hierarchical clustering with the similarity measure $sim_f$. According to \cite{carlssonCharacterizationStabilityConvergence2010}, single-link hierarchical clustering aligns with partitioning by an equivalence relation. This method constructs a hierarchy by initially treating each element as a separate class and merging the closest classes step by step based on similarity. 
A single-link hierarchy evaluates the similarity between classes as the maximum similarity (or minimum dissimilarity) between pairs of elements in the classes.  A classification at similarity threshold $\tau$ merges classes if their similarity exceeds $\tau$. For example, at $\tau = 0.3$, we have two classes: $\{a, b\}$ and $\{c, d, e\}$. When the threshold increases, e.g. $\tau = 0.5$, three classes emerge: $\{a\}$, $\{b\}$, and $\{c, d, e\}$. Further raising the threshold to $\tau = 0.7$ results in four classes: $\{a\}$, $\{b\}$, $\{c, d\}$, and $\{e\}$.

\subsection{Rewriting Trees}
\label{sec:TransTrees}


The aim of the structuring step on line~\ref{Main-ligne:rewrite} of Algorithm~\ref{algo:struct:Main}  is to rewrite the instance tree to conform to a valid schema based on the $\mathbb{G}$ meta-grammar.
Nodes representing \emph{groups}, \emph{relations}, or \emph{collections} are identified, and equivalent sub-trees are rewritten to align with their respective categories.
This iterative process checks schema validity after each modification.
The rewriting approach may vary depending on the objectives.
In this paper, the purpose of the rewriting function is to unify sub-trees by eliminating structural variations, maximizing their frequency, and minimizing the number of grammar production rules.


In Algorithm~\ref{algo:struct:Main}, the evolution from instance $I_i$ to $I_{i+1}$ follows a transformation process, where sub-trees are grouped or modified  based on a similarity measure.
The tree rewriting  function on line~\ref{Main-ligne:rewrite}   takes as input an instance $I_i$, i.e, a single tree representing a rooted forest of instance trees.
Called within the while loop (line~\ref{Main-ligne:while}), this function iteratively transforms the instance into a condensed tree that represents the grammar.  Each iteration focuses on invalid parts of $I_i$, and  when no further transformations are possible, the resulting grammar $G_T$ is valid under $\mathbb{G}$ (end condition of the while loop).
In practice, to ensure termination, a maximum cycle limit $K$ is set, which may leave some tree parts unresolved.

The whole process is governed by four parameters: $f$ (similarity function), $\tau$ (similarity threshold), $minSup$ (minimum element frequency), and $K$ (maximum cycles).

Our rewriting function consists of five main transformation operations, tracked by a variable indicating whether a transformation has occurred. If the tree is modified, further operations are skipped, and the cycle restarts to update equivalence classes. The five \textit{operations}, executed in logical sequence, are: detecting groups, unifying them, grouping into collections, identifying relationships between groups, and relationships between collections.
Operations proceed sequentially; if no changes occur in the earlier steps at iteration $i$, the subsequent operations are applied until a modification occurs, advancing to $I_{i+1}$. If none of the five operations modify the tree, more drastic transformations are applied. 
These final steps work bottom-up, removing intermediate levels above uncategorized entities (operation \textsf{reduce}(bottom)). 
Once all nodes are categorized, the function deletes any remaining upper levels that aren't classified as an \emph{entity}, \emph{group}, \emph{relation}, or \emph{collection} (operation \textsf{reduce}(top)).

The reminder of this section provides an overview of the five key operations employed in the rewriting function  (line~\ref{Main-ligne:rewrite} of Algorithm~\ref{algo:struct:Main}).



\noindent
$\bullet$
The \textit{findGroups} operation identifies frequent groupings of entities in the tree by partitioning the sub-trees above the entity sub-trees.
It filters out these partitions with support below \emph{minSupport}, resulting in a set, denoted \emph{equivalent\_st}, containing sub-tree equivalence classes with sufficient support, as illustrated in the following example.
In general, the \textit{findGroups} operation defines new sub-trees $T_{GROUP}$ with roots labeled $GROUP_k$, where $GROUP$ is a non-terminal symbol and $k$ is an attribute identifying equivalent groupings.
The selection order of equivalence classes in $equivalent\_st$ is determined by the tree depths within each class.

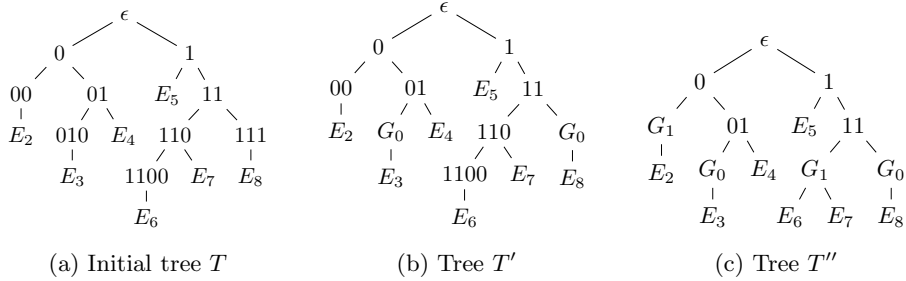
\begin{figure}[htb]
  \centering
  \begin{subfigure}[b]{.3\linewidth}
    \centering
    \begin{adjustbox}{valign=c, max width=\linewidth}
      \begin{forest}
        for tree={s sep=.3em,l sep=.5em,l=1em,fit=tight}
        [$\epsilon$
          [$0$ [$00$ [$E_2$]]
              [$01$ [$010$[$E_3$]]
                  [$E_4$]]]
          [$1$ [$E_5$]
              [$11$ [$110$ [$1100$ [$E_6$]]
                          [$E_7$]]
                  [$111$ [$E_8$]]]]]
      \end{forest}
    \end{adjustbox}
    \caption{Initial tree $T$}
    \label{fig:rewrite:findgroupInit}
  \end{subfigure}
  \hfill
  \begin{subfigure}[b]{.3\linewidth}
    \centering
    \begin{adjustbox}{valign=c, max width=\linewidth}
      \begin{forest}
        for tree={s sep=.3em,l sep=.5em,l=1em,fit=tight}
        [$\epsilon$
          [$0$ [$00$ [$E_2$]]
              [$01$ [$G_0$[$E_3$]]
                  [$E_4$]]]
          [$1$ [$E_5$]
              [$11$ [$110$ [$1100$ [$E_6$]]
                          [$E_7$]]
                  [$G_0$ [$E_8$]]]]]
      \end{forest}
    \end{adjustbox}
    \caption{Tree $T'$}
    \label{fig:rewrite:findgroup1}
  \end{subfigure}
  \hfill
  \begin{subfigure}[b]{.3\linewidth}
    \centering
    \begin{adjustbox}{valign=c, max width=\linewidth}
      \begin{forest}
        for tree={s sep=.3em,l sep=.5em,l=1em,fit=tight}
        [$\epsilon$
          [$0$ [$G_1$ [$E_2$]]
              [$01$ [$G_0$[$E_3$]]
                  [$E_4$]]]
          [$1$ [$E_5$]
              [$11$ [$G_1$ [$E_6$] [$E_7$]]
                  [$G_0$ [$E_8$]]]]]
      \end{forest}
    \end{adjustbox}
    \caption{Tree $T''$}
    \label{fig:rewrite:findgroup2}
  \end{subfigure}

  \caption{Example of the \textsf{findGroups} operation}
  \label{fig:ExFindGroup}
\end{figure}

\begin{example}
  Let $T$ be the tree in Figure~\ref{fig:rewrite:findgroupInit} where entity sub-trees are those with roots labeled $E_i$.
  Let $ D/\equiv_\tau = \{\{0\},$ $\{1\},$ $\{00, 110\},$ $\{01\},$ $\{010, 111\},$ $\{11\},$ $\{1100\}\}$ and $minSupport = 2$.
  Consequently, we have $equivalent\_st = \{\{00, 110\}, \{010, 111\}\}$.
  
  The nodes $010$ and $111$ are relabeled as $GROUP_0$, while the nodes  $00$ and $110$ are assigned the label $GROUP_1$. Node $1100$ is deleted to ensure only entities remain as children of group-nodes, resulting in the updated tree $T''$ (Figure~\ref{fig:rewrite:findgroup2}).\qed.
%
\end{example}

\noindent
$\bullet$ Unifying groups includes to operations: finding sub-groups and merging them.
The \textit{findSubGroups} operation aims at minimizing distinct groups and maximizes their frequencies. It ensures no group contains a more frequent subgroup.
This operation is also applied to sub-trees whose root does not yet have a label corresponding to a non-terminal of the target grammar.
If a more frequent sub-tree $st_i$ can be constructed from a subset of the entity trees descending from a given sub-tree $st$, $st$ is replaced by a new unlabeled sub-tree. The children of this new sub-tree are $st_i$ and the sub-trees in $st$ that are not part of $st_i$, as illustrated in the following example.

\begin{figure}[htb]
    \centering
    \begin{subfigure}{.3\linewidth}
      \centering
      \begin{adjustbox}{valign=c, max width=\linewidth}
        \begin{forest}
          for tree={s sep=.5em,l sep=.5em,l=1em,fit=tight}
            [$\dots$ [$~~$ [$GROUP$ [$E_1$] [$E_2$]] [$E_3$]]]
        \end{forest}
      \end{adjustbox}
    \end{subfigure}
    \hfill
    \begin{subfigure}{.3\linewidth}
      \centering
      \begin{adjustbox}{valign=c, max width=\linewidth}
        \begin{forest}
          for tree={s sep=.5em,l sep=.5em,l=1em,fit=tight}
            [$\dots$ [$~~$ [$GROUP$ [$E_1$] [$E_3$]] [$E_2$]]]
        \end{forest}
      \end{adjustbox}
    \end{subfigure}
    \hfill
    \begin{subfigure}{.3\linewidth}
      \centering
      \begin{adjustbox}{valign=c, max width=\linewidth}
        \begin{forest}
          for tree={s sep=.5em,l sep=.5em,l=1em,fit=tight}
            [$\dots$ [$~~$ [$GROUP$ [$E_2$] [$E_3$]] [$E_1$]]]
        \end{forest}
      \end{adjustbox}
    \end{subfigure}
  \caption{Example of the \textsf{findSubgroups} operation}
  \label{fig:rewrite:findsubgroup}
\end{figure}
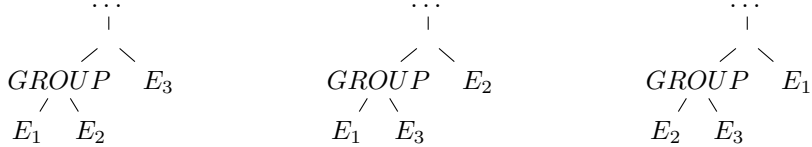

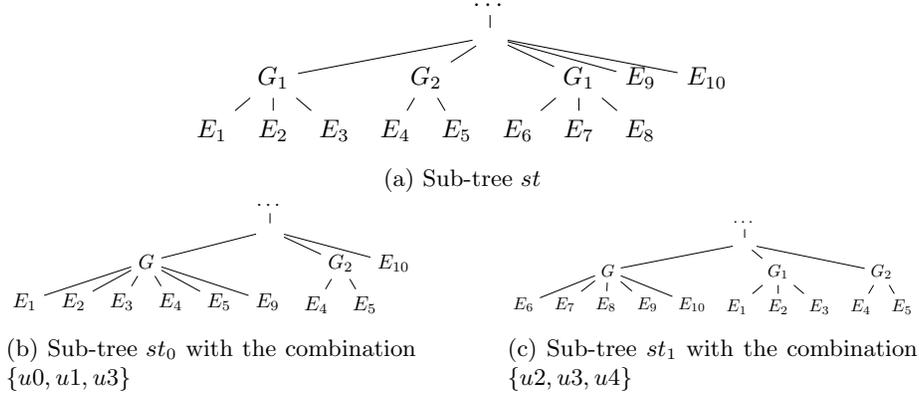
\begin{figure}[htb]
  \centering
  \begin{subfigure}[b]{.6\linewidth}
    \centering
    \begin{adjustbox}{valign=c, max width=\linewidth}
      \begin{forest}
        for tree={s sep=.5em,l sep=.5em,l=1em,fit=tight}
        [$\dots$ [$~~$ [$G_1$ [$E_1$] [$E_2$] [$E_3$]] [$G_2$ [$E_4$] [$E_5$]] [$G_1$[$E_6$] [$E_7$] [$E_8$]] [$E_9$] [$E_{10}$]]]
      \end{forest}
    \end{adjustbox}
    \caption{Sub-tree $st$}
    \label{fig:rewrite:mergegroupsInit}
  \end{subfigure}
  \begin{subfigure}[b]{.45\linewidth}
    \centering
    \begin{adjustbox}{valign=c, max width=\linewidth}
      \begin{forest}
        for tree={s sep=.5em,l sep=.5em,l=1em,fit=tight}
        [$\dots$ [$~~$ [$G$ [$E_1$] [$E_2$] [$E_3$] [$E_4$] [$E_5$] [$E_9$]] [$G_2$ [$E_4$] [$E_5$]] [$E_{10}$]]]
      \end{forest}
    \end{adjustbox}
    \caption{Sub-tree $st_0$ with the combination $\{u0, u1, u3\}$ }
    \label{fig:rewrite:mergegroupsTest1}
  \end{subfigure}
  \hfill
  \begin{subfigure}[b]{.45\linewidth}
    \centering
    \begin{adjustbox}{valign=c, max width=\linewidth}
      \begin{forest}
        for tree={s sep=.5em,l sep=.5em,l=1em,fit=tight}
        [$\dots$ [$~~$ [$G$ [$E_6$] [$E_7$] [$E_8$] [$E_9$] [$E_{10}$]][$G_1$[$E_1$] [$E_2$] [$E_3$]] [$G_2$ [$E_4$] [$E_5$]]]]
      \end{forest}
    \end{adjustbox}
    \caption{Sub-tree $st_1$ with the combination $\{u2, u3, u4\}$}
    \label{fig:rewrite:mergegroupsTest2}
  \end{subfigure}
  \caption{Example of the \textsf{mergeGroups} operation}
  \label{fig:rewrite:mergegroups}
\end{figure}

\begin{example}
Given the tree $T$ in Figure~\ref{fig:rewrite:mergegroupsInit}, let $st$ represent the sub-tree for the first group $G_1$. To check if a subgroup is more frequent than $G_1$, we test all possible subsets of entities. With 3 entity sub-trees in $st$, we test groupings of 2. The results for these groupings are shown in Figure~\ref{fig:rewrite:findsubgroup}.
  \qed
\end{example}

The goal of the \textit{mergeGroups} operation is to increase group size by merging sibling groups or adding sibling entity sub-trees. This operation only applies to groups not already in a relation or collection. For each sub-tree $st$, two sets are considered: $S_{GROUP}$ (children labeled $GROUP$) and $S_{ENT}$ (children labeled $ENT$). The operation explores all combinations between these sets, starting with the largest. Transformations are applied only if they produce more frequent sub-trees than the original. The following  example illustrates this process.

\begin{example}
  Let $T$ be the tree in  Figure~\ref{fig:rewrite:mergegroupsInit}.
  The root of sub-tree $st$ is at position $u$ and is not yet labelled with a target grammar symbol.
  The operation attempts to merge children of $st$. We have $S_{G} = \{u0, u1, u2\}$ and $S_{ENT} = \{u3, u4\}$.
  The number of testable combinations depends on $|S_{G}/\equiv_\tau|$ (Definition~\ref{def:eqClasses}), which is the number of elements in $S_{G}$, excluding those in the same equivalence class. Since $u0$ and $u2$ (both labeled $G_1$) are in the same equivalence class, the \textit{mergeGroups} operation will attempt to build sub-trees by grouping $4$ ($=|S_{G}/\equiv_\tau| +  |S_{ENT}|$) positions as siblings, then $3$ positions, and so on.
  All possible combinations are tested by creating a new tree with the new group and calculating its support.
  With $3$ positions the combinations to be tested are:
  $\{\{u0, u1, u3\},$ $\{u0, u1, u4\},$ $\{u0, u3, u4\},$ $\{u1, u2, u3\},$ $\{u1, u2, u4\},$ $\{u1, u3, u4\},$ $\{u2, u3, u4\}\}$.
  Two of these combinations are shown in Figure~\ref{fig:rewrite:mergegroups}.\qed
\end{example}

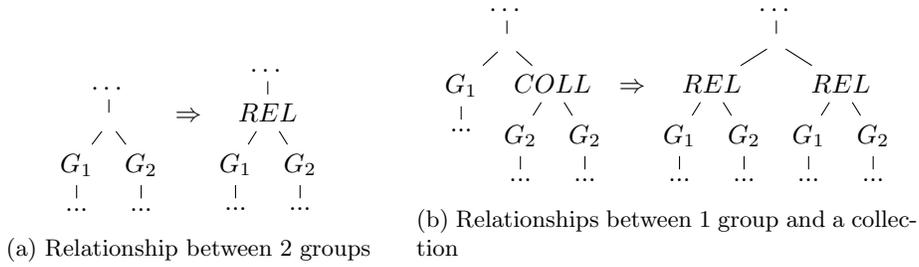
\begin{figure}[htb]
  \centering
  \begin{subfigure}[b]{.4\linewidth}
      \centering
      \begin{adjustbox}{valign=c, max width=\linewidth}
          \begin{forest}
            for tree={s sep=.5em,l sep=.5em,l=1em,fit=tight}
              [$\dots$ [$ $,baseline [$G_1$ [...]] [$G_2$ [...]]]]
          \end{forest}
          $\Rightarrow$
          \begin{forest}
            for tree={s sep=.5em,l sep=.5em,l=1em,fit=tight}
              [$\dots$ [$REL$,baseline [$G_1$ [...]] [$G_2$ [...]]]]
          \end{forest}
      \end{adjustbox}
      \caption{Relationship between 2 groups}
      \label{fig:rewrite:findRelationShipCas1}
  \end{subfigure}
  \hfill
  \begin{subfigure}[b]{.55\linewidth}
      \centering
      \begin{adjustbox}{valign=c, max width=\linewidth}
          \begin{forest}
            for tree={s sep=.5em,l sep=.5em,l=1em,fit=tight}
              [$\dots$ [$ $ [$G_1$,baseline [...]] [$COLL$ [$G_2$ [...]] [$G_2$ [...]] ]]]
          \end{forest}
          $\Rightarrow$
          \begin{forest}
            for tree={s sep=.5em,l sep=.5em,l=1em,fit=tight}
              [$\dots$ [$ $ [$REL$,baseline [$G_1$ [...]] [$G_2$ [...]]] [$REL$[$G_1$ [...]] [$G_2$ [...]]]]]
          \end{forest}
      \end{adjustbox}
      \caption{Relationships between 1 group and a collection}
      \label{fig:rewrite:findRelationShipCas2}
  \end{subfigure}
  \caption{Example of the \textsf{findRelationship} operation}
  \label{fig:rewrite:findRelationShip}
\end{figure}

\noindent
$\bullet$
The operation \textit{findRelations} creates relationships on unlabelled sub-trees in two ways:
\begin{enumerate*}
  \item If the sub-tree has two children labelled $GROUP$, the sub-tree's root is labelled $REL$ (Figure~\ref{fig:rewrite:findRelationShipCas1}).
  \item If the sub-tree has one child  labelled $GROUP$ and another labelled $COLL$, the sub-tree is restructured.
        It creates a new sub-tree with root $REL$ having two children: the child labelled $GROUP$ and another sub-tree derived from the $COLL$ child.
        This transformation is distributive for each child of $COLL$ (Figure~\ref{fig:rewrite:findRelationShipCas2}).
\end{enumerate*}

\begin{figure}[htb]
  \centering
  \begin{adjustbox}{valign=c, max width=\linewidth}
      \begin{forest}
        for tree={s sep=.5em,l sep=.5em,l=1em,fit=tight}
          [$\dots$ [$G_1$,baseline [...]] [$G_2$[...]] [$G_2$[...]] [$G_3$[...]] [$COLL$[$G_1$[...]] [$G_1$[...]]]]
      \end{forest}
      \hspace{1em}
      {\huge$\Longrightarrow$}
      \hspace{1em}
      \begin{forest}
        for tree={s sep=.5em,l sep=.5em,l=1em,fit=tight}
          [$\dots$ [$G_3$,baseline [...]] [$COLL$[$G_1$[...]] [$G_1$[...]] [$G_1$[...]]] [$COLL$[$G_2$[...]] [$G_2$[...]]]]
      \end{forest}
  \end{adjustbox}
  \caption{Example of the \textsf{findCollections} operation}
  \label{fig:rewrite:findCollections}
\end{figure}
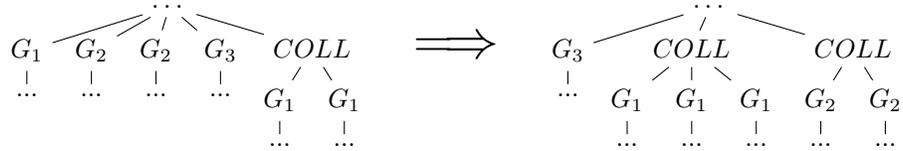

\noindent
$\bullet$
The \textit{findCollections} operation groups sibling sub-trees (either groups or relations) into a collection if they belong to the same equivalence class.
Here we explain how group collections are formed. Relations are handled similarly.
For each \textit{unlabelled} sub-tree $st$, the operation works in three steps (Figure~\ref{fig:rewrite:findCollections}):
\begin{enumerate*}
  \item It creates a subtree labeled $COLL$ with all $GROUP$ subtrees from the same equivalence class as its children;
  \item It groups collections containing children from the same equivalence class;
  \item It adds child groups of $st$ to the collection containing equivalent groups.
\end{enumerate*}

\section{Proof of concept}
\label{sec:POC}
Since approaches in the literature are not directly comparable to ours, we evaluate our prototype based on how well it meets our objectives. We tested our structuring method with a proof-of-concept use case using the CAS corpus~\cite{grabarCASFrenchCorpus2018}, which contains real and fictitious clinical cases describing patients' medical histories, symptoms, diagnoses, and treatments.
We  chose a small example for this initial experiment -- a corpus of $100$ texts and $8098$ manually annotated named entities across $10$ categories, with some entities potentially nested -- to carefully track each step of our method.

\paragraph{Evaluation Methodology.}
We track the instance's evolution by comparing each iteration of Algorithm~\ref{algo:struct:Main}.
The rewriting function (line~\ref{Main-ligne:rewrite}, Section~\ref{sec:TransTrees})   alternates between specializing (making details explicit) and generalizing (aggregating). Due to its behavior, only one operation is triggered per iteration, which can delay annotating nodes as \emph{relation} or \emph{collection}, even when conditions are met. To evaluate progress, we count valid structures at the end of the while loop, even if they aren't fully labelled yet.

The following aspects are guidelines for the analysis of the behavior of our approach:

\begin{enumerate}
  \item \label{asp1} \textit{The impact of our approach on the reduction of production rules.}
  A comprehensible grammar $G_T$ should be more concise than $G_0$, which directly represents the original instance. Reducing the number of production rules reflects aggregation.

  \item \label{asp2} \textit{The average number of instances for core structures defined by $\mathbb{G}$}.
  Our aggregations use \emph{groups}, \emph{relations} and \emph{collections} as the core structures of our schema, and we expect their instances to increase with each iteration.  For example, we track how often a non-terminal name such as $GROUP_9$ appears in the tree. Ideally, you would expect to see a significant number of occurrences.

  \item\label{asp3}  \textit{The hierarchy evolution}.
  Our structures follow a defined hierarchy, with, for instance,  \emph{collections} aggregating \emph{groups} for greater generalization.
  During  the iterations of Algorithm~\ref{algo:struct:Main} we monitor the number of structures in each category: \emph{groups}, \emph{relations}, and \emph{collections}.
\end{enumerate}

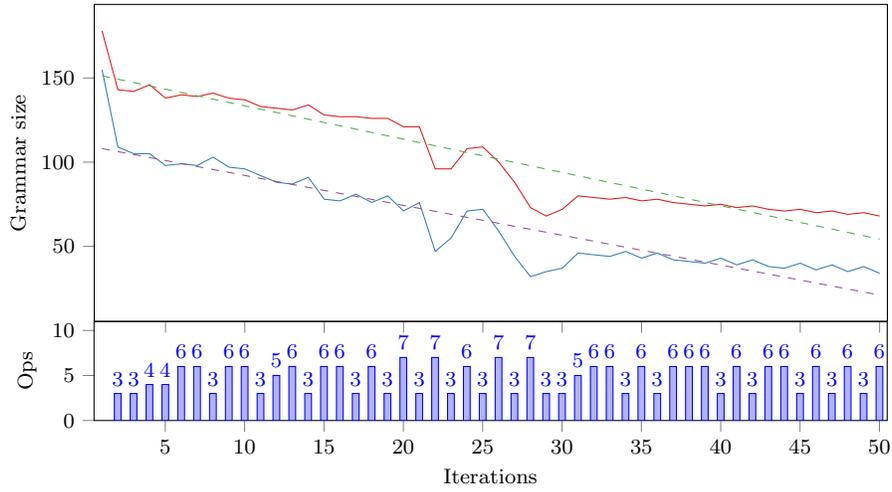
\begin{figure}[htb]
  \centering
  \begin{tikzpicture}
    \footnotesize
    \begin{groupplot}[
        group style={
          group size=1 by 2,
          xlabels at=edge bottom,
          xticklabels at=edge bottom,
          vertical sep=0pt
        },
        enlarge x limits=.01,
        xmin=1,
        xmax=50,
        xlabel={Iterations},
        xtick align=outside,
        ytick align=outside,
        tickpos=left,
        width=\linewidth,
      ]
      \nextgroupplot[ylabel={Grammar size}, height=.3\textheight]]

      \addplot+[cycle list shift=0] table [x expr={1 + \thisrow{step}}, y=value, col sep=comma] {nb_prod.csv}; \label{figure:struct:xp:prod:prod}

      \addplot+[cycle list shift=1] table [x expr={1 + \thisrow{step}}, y=value, col sep=comma] {nb_unlabelled.csv}; \label{figure:struct:xp:prod:unlabelled}

      \addplot+[dashed, cycle list shift=0] table [x expr={1 + \thisrow{step}}, y={create col/linear regression={y=value}}, col sep=comma] {nb_prod.csv}; \label{figure:struct:xp:prod:prod-trend}
      \addplot+[dashed, cycle list shift=1] table [x expr={1 + \thisrow{step}}, y={create col/linear regression={y=value}}, col sep=comma] {nb_unlabelled.csv}; \label{figure:struct:xp:prod:unlabelled-trend}

      \nextgroupplot[ybar, ymin=0, ymax=11, ylabel={Ops}, pattern=crosshatch, height=.15\textheight, bar width=.3em], legend style={ at={(0.5,-0.5)}, anchor=north}

      \addplot+[nodes near coords, cycle list shift=2, fill] table [x=step, y expr={1 + \thisrow{value}}, col sep=comma] {edit_op.csv}; \label{figure:struct:xp:prod:op}
    \end{groupplot}
  \end{tikzpicture}
  \caption{Evolution of the grammar during our approach.}
  \label{figure:struct:xp:prod}
\end{figure}

\paragraph{Experiments.}
Figure~\ref{figure:struct:xp:prod} illustrates the behavior of Algorithm~\ref{algo:struct:Main} in reducing the number of production rules (aspect~\ref{asp1}).
It shows the number of rules in grammar $G_i$ (\ref{figure:struct:xp:prod:prod}) along with its trend line (\ref{figure:struct:xp:prod:prod-trend}), and the number of uncategorized nodes (\ref{figure:struct:xp:prod:unlabelled}) with its trend line (\ref{figure:struct:xp:prod:unlabelled-trend}).
The lower part of the graph shows the transformation applied at each step $i$ (\ref{figure:struct:xp:prod:op}).
Operations number correspond to their order :
\begin{enumerate}
  \item \textsf{findSubgroups};
  \item \textsf{mergeGroups};
  \item \textsf{findCollectionsOfGroups};
  \item \textsf{findRelationships};
  \item \textsf{findCollectionsOfRelationships};
  \item \textsf{reduce}(bottom);
  \item \textsf{reduce}(top).
\end{enumerate}

The algorithm effectively minimizes the grammar, reducing production rules from 178 to 68, and uncategorized nodes from 155 to 34 after 50 iterations.

A detailed analysis of Figure~\ref{figure:struct:xp:prod} reveals moments where the number of production rules remains relatively constant, such as in iterations $21$ and $25$.
During these phases, Algorithm~\ref{algo:struct:Main} applies \emph{generalistion} operations, leading to significant changes in the number of rules. However, these operations do not produce valid or frequent trees, requiring subsequent restructuring. This is where the previously mentioned back-and-forth behavior occurs. In steps $24$, $30$, and $31$, the number of production rules rises sharply as the procedure introduces more specific structures to later recognize more generic ones, like collections.

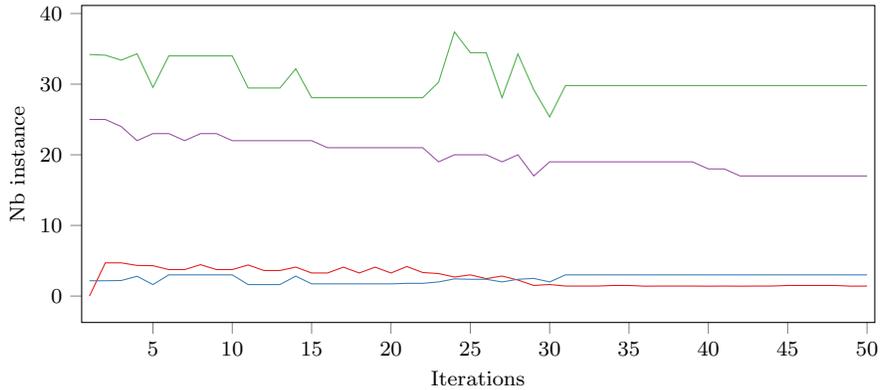
\begin{figure}[htb]
  \centering
  \begin{tikzpicture}
    \footnotesize
    \begin{axis}[
        enlarge x limits=.01,
        xmin=1,
        xmax=50,
        xlabel={Iterations},
        xtick align=outside,
        ytick align=outside,
        tickpos=left,
        legend columns=-1,
        legend style={at={(0.5,-0.25)}, anchor=north},
        width=\linewidth,
        ylabel={Nb instance},
        height=.3\textheight
      ]

      \addplot+ table [x expr={1 + \thisrow{step}}, y=value, col sep=comma] {coll_ratio.csv}; \label{figure:struct:xp:ratio:coll}

      \addplot+ table [x expr={1 + \thisrow{step}}, y=value, col sep=comma] {rel_ratio.csv}; \label{figure:struct:xp:ratio:rel}

      \addplot+ table [x expr={1 + \thisrow{step}}, y=value, col sep=comma] {group_ratio.csv}; \label{figure:struct:xp:ratio:group}

      \addplot+ table [x expr={1 + \thisrow{step}}, y=value, col sep=comma] {nb_equiv_subtrees.csv}; \label{figure:struct:xp:ratio:equiv}
    \end{axis}
  \end{tikzpicture}
  \caption{Mean number of instance for each structure}
  \label{figure:struct:xp:ratio}
\end{figure}

Figure~\ref{figure:struct:xp:ratio} shows the behavior of Algorithm~\ref{algo:struct:Main}  regarding the number of instances for each structure (aspect~\ref{asp2}). The curve~\ref{figure:struct:xp:ratio:equiv} represents the number of equivalence classes at each step, which decreases from $25$ to $17$.  This suggests that certain structures unify over time, with the transformations successfully remodeling different groups that eventually fall into the same equivalence class.
The curves~\ref{figure:struct:xp:ratio:group}, \ref{figure:struct:xp:ratio:rel}, and \ref{figure:struct:xp:ratio:coll} show the average number of instances for each \emph{group}, \emph{relation}, and \emph{collection}, respectively. Despite the decrease in unlabeled nodes, these numbers remain relatively stable. Initially, the average number of instances is $34.2$ for groups and $2.2$ for relations. By the end, groups have an average of $29.8$ instances and relations $3$. The drop in group instances is due to their promotion to relations, increasing the average for relations. The decrease in collection instances is explained by successive merging, resulting in $1.4$ instances per collection after $50$ iterations. Ideally, there should be $1$ collection per \emph{group} and per \emph{relation}, meaning all collections have been successfully merged.

\begin{figure}[htb]
  \centering
  \begin{tikzpicture}
    \footnotesize
    \begin{axis}[
        enlarge x limits=.01,
        xmin=1,
        xmax=50,
        xlabel={Iterations},
        xtick align=outside,
        ytick align=outside,
        tickpos=left,
        legend columns=-1,
        legend style={at={(0.5,-0.25)}, anchor=north},
        width=\linewidth,
        ylabel={Nb structures},
        height=.3\textheight
      ]

      \addplot+ table [x expr={1 + \thisrow{step}}, y=value, col sep=comma] {nb_coll.csv}; \label{figure:struct:xp:nbElems:coll}

      \addplot+ table [x expr={1 + \thisrow{step}}, y=value, col sep=comma] {nb_rel.csv}; \label{figure:struct:xp:nbElems:rel}

      \addplot+ table [x expr={1 + \thisrow{step}}, y=value, col sep=comma] {nb_group.csv}; \label{figure:struct:xp:nbElems:group}
    \end{axis}
  \end{tikzpicture}
  \caption{Number of each kind of structures}
  \label{figure:struct:xp:nbElems}
\end{figure}
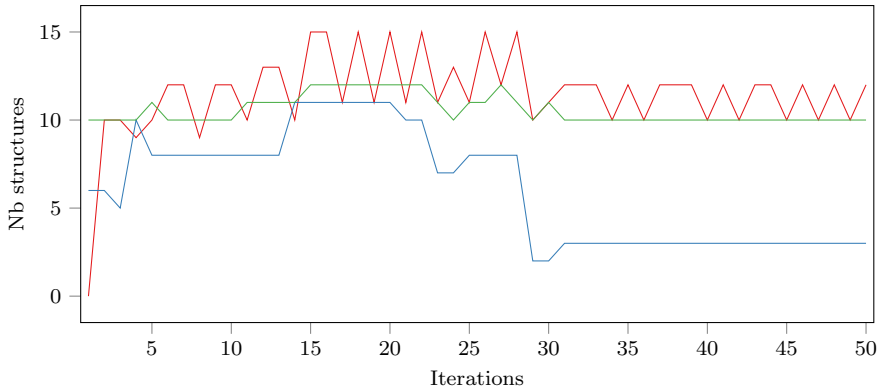

Figure~\ref{figure:struct:xp:nbElems} illustrates the behavior of Algorithm~\ref{algo:struct:Main} regarding the number of structures per category (aspect~\ref{asp3}). Specifically, \ref{figure:struct:xp:nbElems:group} shows the number of groups, curve~\ref{figure:struct:xp:nbElems:rel} the relations, and \ref{figure:struct:xp:nbElems:coll} the collections.
In instance $I_0$ (iteration 1), $10$ groups, $6$ relations, and $0$ collections are formed. By iteration $28$, the procedure creates $11$ groups, $8$ relations, and 15 collections, with the fewest production rules. Iterations $29$ and $30$  merge collections after the tree's top reduction in iteration $28 $(see Section~\ref{sec:TransTrees}). This reduces relations to $2$, with a third added at iteration $31$, due to the reduction in groups from $15 $ to $10$. Fewer groups lead to fewer relations, but more instances per relation as equivalence classes merge. A similar unification occurs at iteration $21$.
The procedure concludes with $10$  groups, $3$ relations, and $12$ collections (nearly one collection per group and relation).
After iteration $31$, the number of groups and collections stabilizes, indicating the structure effectively represents the data. Subsequent steps only modify or refine the instance without altering the schema significantly.

\noindent
\paragraph{Reproducibility.}
The procedures are described in detail in \cite{hiotConstructionAutomatiqueBases2024}.
The implementation is written in Python and utilizes CoreNLP for parsing.
The source code, required to reproduce the experiments, can be found on GitHub~\cite{chabin\architxt2024}.
For the corpus, \cite{grabarCASFrenchCorpus2018} provides the data access modalities.

\noindent
\paragraph{Evaluating the rewriting policy.}
The policy of Algorithm~\ref{algo:struct:Main} for identifying frequent elements converges to a satisfactory solution but is highly sensitive to parameters like $ minSup$ or $\tau$. For example,  consider changing $\tau$ from $0.5$ to $0.7$ on a tree where a node has $X$- rooted  and $Y$-rooted sub-trees as children, both containing $GROUP_7$ and \emph{Anatomy} as their children.
We might expect the \textit{mergeGroups} operation, mentioned in Section~\ref{sec:TransTrees}, to merge $GROUP_7$ with \emph{Anatomy}. However, this merges fails. Adding \emph{Anatomy} to $GROUP_7$ places the group too far from other instances, creating a set with insufficient support. This happens because transformations are made individually, not collectively; simultaneous modifications could have enabled the merge.

\noindent
\paragraph{Evaluating  the resulting grammar.}
The target grammar $G_T$, obtained from our experiments on the CAS corpus, is valid with respect to $\mathbb{G}$ and consists of 26 production rules (12 collections, 3 relations, 10 groups, and 9 entities).
Analyzing the groups and their semantics reveals expected associations, such as:
$GROUP_0 \to ENT_{Dose}$ $ENT_{Frequence}$ $ENT_{Mode}$ $ENT_{Substance}$ $ENT_{Sosy}$\\$ENT_{Treatment}$ indicating a treatment;
$GROUP_1 \to ENT_{Exam}$ $ENT_{Value}$;
and $GROUP_3 \to ENT_{Dose}$ $ENT_{Exam}$ $ENT_{Sosy}$\\$ENT_{Substance}$ indicating different formats for exam results.

The results are very promising: the obtained generic schema is coherent and reliable.
 While differences with human analysis are noted and will be explored in the following discussion, it's important to understand that our method relies primarily on the frequency of syntactic structures, with semantic information mainly derived from the enrichment step. Consequently, although the results are coherent and logical, they do not  always fit perfectly with an intuitive database model. Here are some examples.
\begin{enumerate}
    \item $G_T$ includes the production rules:
    $GROUP_4 \to ENT_{Anatomy},$ $ENT_{Exam},$ $ENT_{Sosy}$ and
    $GROUP_8 \to ENT_{Exam},$ $ENT_{Sosy}$.
    While domain knowledge may suggest splitting them into two groups — one for exams (\emph{Exam}, \emph{Anatomy}) and one for symptoms (\emph{Sosy}, \emph{Anatomy}) — connected by a relation, the corpus shows these entities are often interdependent with few external links.
    Grouping them together is therefore consistent, reducing grammar complexity by avoiding extra groups and relations.

    \item A prescription typically includes elements like \emph{Treatment}, \emph{Substance}, \emph{Dose}, \emph{Mode}, and \emph{Frequency}, related to a symptom (\emph{Sosy}).
    In the corpus, treatments are almost always tied to symptoms, naturally forming $GROUP_0$ with both.
    While a database model might separate the symptom into its own group to connect it to both exams and treatments, our approach doesn't account for such domain-specific structures.
    Instead, it combines them (e.g., production rules for $GROUP_0$, $GROUP_4$ and  $GROUP_8$ above), reducing grammar size by avoiding extra relations.
    If the corpus had more isolated instances of \emph{Sosy}, our method might have created a distinct \emph{Sosy} group, leading to more relations.

    \item $G_T$ provides various production rules to represent examination results, such as $GROUP_3 \to ENT_{Dose}$ $ENT_{Examination}$
    $ENT_{Sosy}$ $ENT_{Substance}$ and those for groups 1, 4 and 8 (mentioned above).
    Although this can be criticized, each group reflects different types of examination: some are only mentioned (group 4), some result in a value (group 1) and others measure a dose (group 3).
    In particular, $GROUP_3$ is ambiguous and may also represent post-test treatments.
    As in the previous point, the presence of \emph{Sosy} could also be handled by a relation.
    
    \item For our experiments, the resulting $G_T$ offers few production rules for relations.
    In some cases we might question whether a relation should be classified as a group or vice versa.
    Indeed, our approach may decide to define a relation between groups representing entities with many missing instances, rather than grouping these entities together.
    This highlights that, in this case, while the semantics may be similar, the similarity of the corresponding sub-trees is quite low.
\end{enumerate}

The resulting $G_T$ defines a general database structure, which can be used to implement various database models. Our approach (and this paper) does not cover the next step - selecting a specific database model based on $G_T$. This step should account for semantic and performance factors, as discussed in Section~\ref{sec:RelW}. In our example, a database analyst could convert $G_T$ into a relational schema with $5$ tables: 
$Prescription[treatment,$ $mode,$ $substance,$ $dose,$ $frequency$, $sosy]$ 
(describes the prescription of a substance or treatment for a symptom);
$Examination[examId, exam ,$ $value]$ and $Measure[examId, exam,$ $dose,$ $substance]$  
(represent two ty\-pes of exams: basic one the those measuring substance levels e.g., in blood);
$forAnat[examId,$ $anatomy]$ (linking exams to part of the body) and
$forSosy[examId,$ $sosy]$  (linking exams to signs or symptoms).
No table links examinations/symptoms to treatments - they typically appear in different sentences and the system is limited to sentence-level associations.



\section{Concluding Remarks}
\label{sec:conclusion}

Our approach aims to organize unstructured data into a flexible structure that abstracts different database models. 
Currently, the resulting grammar $G_T$ can be used as input for other methods that focus on semantic and performance aspects to propose a specific database model.
The weakness of our method, noted in Section~\ref{sec:RelW}, stems  from the use of limited semantic information.
To address this, enhancements could include incorporating semantics through tree rewriting strategies or within the meta-grammar $\mathbb{G}$, taking into account  business rules and  functional dependencies. 
Additionally, integrating functional dependency discovery, as explored in \cite{papenbrockFunctionalDependencyDiscovery2015}, could further refine the structuring process through semantic analysis
Future work also includes extending the proof of concept, improving performance, and exploring incremental structuring methods.

The strength of our approach lies in discovering a schema while structuring the data, making it easier to create database instances from textual data. Avoiding targeting one precise database model we incorporates a multi-model philosophy.
Our method stands apart from both traditional and machine learning information extraction methods by offering a hybrid strategy that provides some flexibility across application domains without the need for training data.
Its transparent process allows users to track and validate each step, ensuring confidence in both the system and the quality of the data.

Finally, our paper can be seen as an instantiation of a  generic and  innovative 
 approach to structuring textual data using tree rewriting and grammar extractions guided by a meta-model defined by an attribute grammar $\mathbb{G}$.  Indeed,  the idea here is to structure  textual data into a format that conforms to a predefined framework, specified by a meta-model $\mathbb{G}$, which  can specify different target structures.
  The textual data is represented as a rooted forest based on an initial grammar, $G_0$.
The process involves transforming this initial structure to meet the constraints of a new format defined by a target grammar, $G_T$. 
The meta-model $\mathbb{G}$ 
guides the process by outlining the desired structure, with transformations applied to achieve it. 
These transformations occur incrementally, evolving both the data and the grammar from $G_0$ to $G_T$.

%

\section{Acknowledgments}
We acknowledge the support of the APR-IA DOING project and the DATA project from ARD-JUNON, both funded by the Centre Val de Loire region in France.

\bibliographystyle{plainurl}
\bibliography{biblio}

\end{document}